\documentclass{ar-1col-S2O}
\usepackage[numbers]{natbib}
\usepackage{url}

\usepackage{float}

\jname{Xxxx. Xxx. Xxx. Xxx.}
\jvol{AA}
\jyear{YYYY}
\doi{10.1146/((please add article doi))}

\def \lesssim {\mathrel{\vcenter
     {\offinterlineskip \hbox{$<$}\hbox{$\sim$}}}}
\def \gtrsim {\mathrel{\vcenter
     {\offinterlineskip \hbox{$>$}\hbox{$\sim$}}}}

\begin{document}

\markboth{Hans-Thomas Janka}{Long-Term Supernova Models}

\title{Long-Term Multidimensional Models of Core-Collapse Supernovae: Progress and Challenges}

\author{Hans-Thomas Janka$^1$
\affil{$^1$Max Planck Institute for Astrophysics, Garching, Germany, Karl-Schwarzschild-Str.~1; email: thj@mpa-garching.mpg.de}}


\begin{abstract}
Self-consistent, multidimensional core-collapse supernova (SN) simulations, especially in 3D, have achieved tremendous progress over the past 10 years. They are now able to follow the entire evolution from core collapse through bounce, neutrino-triggered shock revival, shock breakout at the stellar surface to the electromagnetic SN outburst and the subsequent SN remnant phase. Thus they provide general support for the neutrino-driven explosion mechanism by reproducing observed SN energies, neutron-star (NS) kicks, and diagnostically relevant radioactive isotope yields; they allow to predict neutrino and gravitational-wave signals for many seconds of proto-NS cooling; they confirm correlations between explosion and progenitor or remnant properties already expected from previous spherically symmetric (1D) and 2D models; and they carve out various scenarios for stellar-mass black-hole (BH) formation. Despite these successes it is currently unclear which stars explode or form BHs, because different modeling approaches disagree and suggest the possible importance of the 3D nature of the progenitors and of magnetic fields. The role of neutrino flavor conversion in SN cores still needs to be better understood, the nuclear equation of state including potential phase transitions implies major uncertainties, the SN~1987A neutrino measurements raise new puzzles, and tracing a possible correlation of NS spins and kicks requires still more refined SN simulations.
\end{abstract}

\maketitle

\tableofcontents

\section{INTRODUCTION} 

Supernova~1987A had a pivotal influence on supernova (SN) research, on astrophysics in general, and on astro-particle physics in particular, because it marked the advent of multimessenger astronomy by the historical detection of the first extragalactic neutrinos as unique benefit in addition to a huge wealth of observational data in a wide range of electromagnetic wavelengths (e.g., \citenum{Arnett+1989,Hillebrandt+1989,Danziger+1991,Nomoto+1994}). The two dozen neutrinos \citep{Hirata+1987,Bionta+1987,Alekseev+1988} signaled the existence, at least transiently, of a neutron star (NS), which had formed by the core collapse (CC) of the first blue supergiant, Sanduleak --69$^\circ$202, that was caught to explode as an unusual SN of Type~IIP, i.e., with the ejection of a massive hydrogen-helium envelope. 

Numerous observational features (for a summarizing overview, see, e.g., \citenum{Utrobin+2021}, and references therein) indicated that the SN ejecta were asymmetric and clumpy, for example the broad and dome-shaped peak of the light curve, Doppler shifts and time-evolving substructures of spectral lines, and the unexpectedly early detection of X-ray and $\gamma$-ray emission from radioactive decays. Moreover, radioactive isotopes, in particular $^{56}$Ni, had been mixed from the site of their creation close to the proto-neutron star (PNS) far into the hydrogen envelope of the exploding star. 

Large-scale mixing due to Rayleigh-Taylor instability at the composition-shell interfaces (mainly the He/H interface) after the passage of the outward propagating shock had been expected in stellar explosions and had been witnessed in early multidimensional (MD) simulations, which were mostly constrained to two spatial dimensions (2D, i.e., axisymmetry; \citenum{Chevalier+1978,Arnett+1989b,Ebisuzaki+1989,Benz+1990,Fryxell+1991,Mueller+1991,Herant+1991,Herant+1992}).
\begin{marginnote}[]
\entry{MD}{Multidimensional, i.e., in two spatial dimensions (axisymmetric; 2D) or three-dimensional (3D)}
\end{marginnote}
However, it quickly became clear that shell mixing in the outer stellar layers could not explain the high spectral-line velocities of iron-group elements and the discovery of X-rays and $\gamma$-rays already a few months after the first sighting of SN~1987A \citep{Dotani+1987,Itoh+1987,Sunyaev+1987,Leising1988}. Therefore the focus shifted to the collapsed stellar core and the early post-bounce evolution, i.e., to the region and time where and when the radioactive nuclei are expected to be created. Indeed, 2D hydrodynamic simulations with an approximate treatment of the equation-of-state (EoS) of the stellar plasma and simplified neutrino heating and cooling demonstrated a rapid development of strong convective overturn in the gain layer between the ``gain radius''~\citep{Bethe+1985} and the stalled bounce shock, where the absorption of electron neutrinos ($\nu_e$) and antineutrinos ($\bar\nu_e$) by free neutrons and protons, respectively, dominates the inverse neutrino-cooling processes. Therefore energy is deposited and creates a negative entropy gradient \citep{Herant+1992b}, which triggers non-radial mass motions that imprint large-scale asymmetries on the developing explosion already during the very first second of the neutrino-powered SN blast.

Subsequent 2D simulations with gray (i.e., energy-integrated) flux-limited diffusion (FLD) for neutrinos and a more detailed treatment of the EoS and neutrino rates \citep{Herant+1994,Burrows+1995} as well as a systematic investigation of the neutrino-heating conditions~\citep{Janka+1995,Janka+1996} showed that postshock convection lowers the critical neutrino luminosity threshold for reviving the stalled SN shock compared to spherically symmetric (1D) simulations \citep{Janka1993,Janka+1993,Burrows+1993}. Neutrino-heated matter rises in buoyant plumes, pushes the stagnant shock to larger radii, and thus escapes from energy losses through the re-emission of neutrinos in the cooling layer between gain radius and PNS. The buoyant plumes also give way to accretion downflows that carry fresh, low-entropy matter close to the gain radius, where the gas readily absorbs energy from neutrinos. Thus the mass and volume of the gain layer increase and energy transfer by neutrinos becomes more efficient. Due to this positive feedback, a runaway expansion of the SN shock can set in for lower neutrino luminosities than in 1D. 

The effects of MD mass motions in the gain layer have been subsumed as ``convective engine''~\citep{Herant+1994} or as ``postshock turbulence''~\citep{Burrows+1995,Murphy+2013,Abdikamalov+2015,Couch+2015,Radice+2015,Radice+2016,Radice+2018}. However, both concepts cannot adequately capture the full complexity of the non-linear, highly time-dependent behavior of the postshock flows \citep{Mueller+2015,Mueller2016,Janka+2016}. For example, rather than being determined by small-scale turbulent eddies, the shock evolution is mainly affected by large-scale asymmetries manifesting themselves in the biggest buoyant plumes, bubbles, and vortices, which carry most of the thermal and kinetic energy in the gain layer. This fact is corroborated by the impact of the standing accretion shock instability (SASI) in its sloshing and spiral versions \citep{Blondin+2003,Blondin+2007}. SASI is a global, non-radial instability of the stalled accretion shock that grows in an oscillatory manner by an advective-acoustic cycle~\citep{Foglizzo+2007,Foglizzo2024} with highest growth rates in the dipole and quadrupole modes \citep{Ohnishi+2006,Iwakami+2009,Fernandez+2010,Fernandez+2014,Buellet+2023}. When SASI shock motions meet favorable growth conditions, which is enabled by small shock-stagnation radii and phases of shock retraction~\citep{Scheck+2008,Burrows+2012,Mueller+2012b,Fernandez+2014}, SASI activity leads to secondary postshock convection, tends to support shock expansion, and facilitates more efficient neutrino-energy transfer to the shock \citep{Scheck+2008,Marek+2009}.
\begin{marginnote}[]
\entry{SASI}{Standing Accretion Shock Instability}
\end{marginnote}

MD hydrodynamic instabilities in the deep interior of SNe are therefore crucial for the success of the neutrino-driven explosion mechanism. Yet, the viability of this mechanism and its ability to explain observed SN energies still depends on sufficiently strong energy transfer by the neutrinos~\citep{Janka+1996,Mezzacappa+1998}. Therefore self-consistent, first-principle hydrodynamic simulations with elaborate neutrino transport are indispensable to investigate this problem. Such simulations start from progenitor structures at the pre-collapse stage (or shortly earlier), obtained from stellar evolution calculations, and they follow the collapse of the degenerate cores of the stars by solving the neutrino-hydrodynamics equations with state-of-the-art microphysics, i.e., neutrino reaction rates, EoS of the stellar plasma, and a treatment of nuclear composition changes by, in the best case, a nuclear reaction network. The goals of such calculations are to decide, based on currently best knowledge, whether the stars explode or not, and to determine the properties of the SNe and of the compact remnants of stellar CC, either neutron stars (NS) or black holes (BHs).  

A great number of increasingly sophisticated simulations of this kind in 2D and three dimensions (3D) over the past 30 years has shed light on the intricate coupling of neutrino processes and hydrodynamics in SN cores. This work has not only demonstrated the basic functioning of the neutrino-driven mechanism on grounds of our growing understanding of the involved physics, but it has also gradually improved the theoretical predictions of observable multimessenger signals from CCSNe: neutrinos, gravitational waves (GWs), and electromagnetic radiation. Moreover, progress in this area has begun to bridge the gap between stellar progenitors and their explosions by MD modeling on both sides, and it has moved forward to connect the first-principle models of the central ``engine'' to observational properties of CCSN explosions and their gaseous and compact remnants. In this context, 3D simulations and, in particular, long-term calculations that cover the evolution from the late pre-collapse stage of the progenitors through CC, bounce, and shock revival toward shock breakout from the star and beyond, play a pivotal role for a detailed and quantitative comparison of models and observations. Numerous reviews and book chapters report on the achievements in this long-lasting enterprise \citep{Janka+2007,Janka2012,Janka+2012,Janka+2016,Janka2017,Mueller2016,Mueller2020, Burrows+2021,Yamada+2024}, and a brief survey from the historical perspective will be provided in Section~\ref{sec:MDhistory}.

This review will especially focus on the recent progress in performing long-term MD simulations of hydrodynamical and particle processes that take place at the center of the SN after the onset of the explosion. These simulations are now able to follow the SN blast until the explosion energy saturates, until the nucleosynthesis by freeze-out from nuclear statistical equilibrium (NSE) and explosive nuclear burning is finished, and until some of the birth properties of the compact remnants are defined. The relevant period of time extends over many seconds to well beyond 10\,s, which has been out of reach of self-consistent, first-principle MD models including neutrino physics until recently. The extreme numerical challenges and computational demands, in particular in 3D, have become manageable now by dint of more efficient and robust codes, huge amounts of resources on massively parallel supercomputers, and smart strategies to save such resources by suitable approximate treatments. Traversing this period of time when neutrino physics is crucial with self-consistent models is also of great importance to overcome the limitations of existing 3D simulations that connect neutrino-driven explosions with SN observations by studying mixing and asymmetries in SN explosions, SN light curves, and the morphological properties of SN remnants such as those of the Cassiopeia~A (Cas~A) remnant revealed by the stunningly detailed survey of the James Webb Space Telescope~\citep{Milisavljevic+2024}. Although these 3D simulations traced the shock propagation from core bounce to breakout from the stellar surface (minutes to days later, depending on the star's radius), and farther through the SN phase (some years) into the SN remnant stage (hundreds of years and longer), they have been performed so far with a parametric treatment of neutrino heating to initiate and power the SN blast wave until the explosion energy has reached its terminal value (e.g., \citenum{Hammer+2010,Wongwathanarat+2013,Wongwathanarat+2015,Wongwathanarat+2017,  Utrobin+2021,Gabler+2021,Orlando+2021}).

This review will highlight new insights obtained from long-term simulations of how CCSNe acquire their final energies, how BHs can form in different kinds of scenarios with and without accompanying SN explosion, how asymmetric mass ejection and anisotropic neutrino emission cooperate in kicking new-born NSs and BHs, how 3D effects modify radiated neutrino and gravitational-wave signals, and how the conditions for iron-group and trans-iron nucleosynthesis are affected by the MD flow dynamics of the neutrino-heated, innermost SN ejecta. Within the limitations of this article the primary emphasis will be on neutrino-driven explosions, which are likely to explain the far majority of SNe with energies up to (2--3)\,B, but magnetohydrodynamic effects will be touched if relevant in the context.
\begin{marginnote}[]
\entry{Bethe}{$1\,\mathrm{B} = 1\,\mathrm{bethe} = 10^{51}\,\mathrm{erg} = 10^{44}\,\mathrm{J}$}
\end{marginnote}
The role of nuclear and particle physics inputs for the simulations will be addressed, also in their relevance for observables such as neutrinos and heavy-element formation. Although the main focus will be on long-term SN simulations, the discussion will also include the still unsettled question of the ``explodability'' of massive stars, i.e., which progenitors succeed and which ones fail to explode. Moreover, despite the main weight will be on MD simulations, the possibilities as well as limitations and deficiencies of 1D models and their differences compared to MD results will be addressed, too.

\begin{textbox}[h]\section{SUGGESTIONS FOR ALTERNATIVES TO THE NEUTRINO-DRIVEN MECHANISM}
The {\bf magnetorotational mechanism} taps the reservoir of rotational energy of a rapidly spinning PNS or of a BH with a thick accretion torus via ultra-strong magnetic fields to expel mass in (possibly relativistic) jets and a stellar explosion. It is a viable alternative (or supplement) to neutrino-driven explosions for fast-rotating progenitors and is considered as an explanation of long-duration gamma-ray bursts, hypernovae, some superluminous SNe, and possibly stripped-envelope SNe that possess unusually high energies and signatures of axis-dependent asymmetries or jets \citep[e.g.,][]{Woosley+2006,Bugli+2021,Obergaulinger+2022,Powell+2023}. 
A {\bf hadron-quark phase transition} might release considerable additional amounts of gravitational binding energy and could thus trigger a SN explosion by a second shock wave and/or enhanced neutrino heating~\citep{Fischer+2018,Jakobus+2022}. The conditions for this to happen without conflicting with well-established nuclear theory and astronomical constraints require fine tuning and need to be substantiated. 
The so-called {\bf ``jittering-jets'' explosion mechanism} (\citenum{Soker+2024} and references therein) assumes short-lived accretion disks with time-variable orientation to release energy in jet outflows, thus triggering the SN explosion. This scenario is not solidified by first-principle, quantitative MD simulations and its underlying conceptual assumptions are not supported by current self-consistent models. Instead, the scenario is mainly motivated by a special interpretation of specific morphological features of CCSN remnants (\citenum{Bear+2025} and references therein), whose origin may well be explained by neutrino-driven explosions or could be connected to asymmetries in the circumstellar environment rather than intrinsic processes at the center and the beginning of the explosions.
A {\bf thermonuclear mechanism} assumes that energy released from explosive nuclear burning of helium, which was mixed into the deeper carbon and oxygen layers of rotating progenitor stars and ignites during stellar collapse, could eject the overlying stellar matter~\citep{Kushnir2015,Blum+2016}. The pre-collapse models for this scenario are artificially constructed and their internal structure and composition are not compatible with self-consistent stellar evolution calculations. Moreover, it is questionable that such explosions would be compatible with observational constraints on NS masses and SN nucleosynthesis.
\end{textbox}

\section{NEUTRINO-DRIVEN EXPLOSIONS IN MULTIDIMENSIONAL MODELS}
\label{sec:Nu-Explosions}

\subsection{Brief Historical Survey: The Route to 3D Supernova Modeling}
\label{sec:MDhistory}

\subsubsection{Developments in 2D Paving the Way}

Self-consistent simulations of neutrino-driven CCSNe in 3D were pioneered by Fryer \& Warren~\citep{Fryer+2002,Fryer+2004}, following the modeling approach of Reference~\citep{Herant+1994} by using still relatively low-resolution smoothed-particle hydrodynamics, gray neutrino diffusion, and tracing the neutrino-triggered shock expansion for rather short periods of time. Contemporaneously, more sophisticated multigroup (i.e., energy-dependent with a binned energy spectrum) neutrino transport methods were developed for 2D (and 1D) neutrino-hydrodynamics simulations. On the one hand the Newtonian \textsc{Vulcan} code~\citep{Livne+2004} employed a multi-angle (S$_\mathrm{n}$) Boltzmann solver or a FLD approximation, however did not take into account energy-bin coupling and terms that depend on the velocity of the stellar plasma. On the other hand, the \textsc{Prometheus-Vertex} code \citep{Rampp+2002,Buras+2006,Marek+2006} coupled the Newtonian hydro code \textsc{Prometheus} \citep{Fryxell+1991,Mueller+1991} with the \textsc{Vertex} transport scheme, which integrates the two-moment (i.e., neutrino energy and momentum) equations with a variable Eddington closure obtained from a consistent solution of the Boltzmann equation, including energy-bin coupling, velocity-dependent (${\cal O}(v/c)$) terms, and general relativistic (GR) corrections in the gravitational potential and transport, but employing the so-called ray-by-ray-plus (RbR+) approximation for MD simulations (more details in Section~\ref{sec:PVNmethods}).
\begin{marginnote}[]
\entry{RbR(+)}{Ray-by-ray(-plus) transport approximation}
\end{marginnote}

\begin{marginnote}[]
\entry{AEF}{Algebraic Eddington factor or tensor to close the two-moment equations for neutrino energy and momentum}
\end{marginnote}
Similar developments followed later with a multigroup MD FLD treatment in the Newtonian neutrino-hydrodynamics code \textsc{Castro}~\cite{Zhang+2013}, a GR multigroup RbR+ FLD neutrino transport applied with Newtonian hydrodynamics in the \textsc{Chimera} code~\cite{Bruenn+2020}, the GR hydrodynamics \textsc{CoCoNuT} code with \textsc{Vertex} RbR+ neutrino transport~\citep{Mueller+2010} and thereafter with ``fast multigroup transport'' (FMT) with RbR treatment (\citenum{Mueller+2015}; for stationary solutions, ignoring energy bin coupling and velocity terms, employing a simplified closure and a reduced set of neutrino rates with approximate descriptions), the so-called isotropic diffusion source approximation (IDSA) scheme of Reference~\citep{Liebendoerfer+2009} in a RbR approach in the \textsc{Zeus}-IDSA and 3DnSNe-IDSA codes, the latter more recently upgraded by GR corrections~\citep{Suwa+2013,Takiwaki+2014,Kotake+2018}, the 3D GR \textsc{Zelmani} code with two-moment neutrino transport using an analytic closure via an algebraic Eddington factor (AEF) but ignoring velocity dependence and inelastic scattering processes (\citenum{Roberts+2016}; see also Reference~\citenum{Kuroda+2016} for a similar scheme including velocity terms in the transport), the GR \textsc{Nada-FLD} hydrodynamics code with multigroup MD FLD~\citep{Rahman+2019}, the Newtonian hydrodynamics code \textsc{Alcar} with MD, energy- and velocity-dependent two-moment neutrino transport with AEF closure and GR corrections in the gravitational potential and transport equations~\citep{Just+2015,Just+2018}, and similar developments with the \textsc{Flash}~\citep{OConnor+2018b,OConnor+2018} and \textsc{Fornax} codes~\citep{Burrows+2018,Skinner+2019}. An even more challenging and computationally more demanding code development strove for coupling MD hydrodynamics with the MD Boltzmann transport, also in GR~\citep{Nagakura+2014,Nagakura+2017,Akaho+2021}.

This listing is probably not complete; for more information the reader is referred to References~\citep{OConnor+2018c,Mezzacappa+2020}. Nevertheless, the list demonstrates the enormous body of work that has been invested on the theory and coding in order to advance first-principle CCSN modeling. It also shows that corresponding CCSN simulations differ in the applied transport schemes including the implementation of the neutrino reactions, but they also differ with respect to the numerical integration methods of the hydrodynamics equations and the employed computational grids, i.e., coordinate systems, special grid configurations such as Yin-Yang patches~\citep{Kageyama+2004,Wongwathanarat+2010}, refinement or coarsening procedures, and mesh resolution in different domains or coordinate directions. All of these aspects have advantages and disadvantages as well as specific properties of numerical accuracy and computational performance. Such issues can be important for interpreting the simulation results, for judging the reliability of conclusions about the explosion mechanism, and for understanding the limitations of simulations that connect first-principle SN models to observationally relevant phenomena (for additional reviews, see \citenum{Janka+2016,Mueller2020}).  

2D simulations with \textsc{Prometheus-Vertex} \citep{Buras+2006b,Marek+2009,Janka2012,Summa+2016,Bollig+2017}, \textsc{Chimera} \citep{Bruenn+2013,Bruenn+2016,Bruenn+2023}, \textsc{CoCoNuT-Vertex}~\citep{Mueller+2012a,Mueller+2012b,Janka+2012}, the 3DnSNe-IDSA  method~\citep{Nakamura+2015}, \textsc{Zeus}-IDSA~\citep{Suwa+2016}, \textsc{Flash}~\citep{OConnor+2018b}, and \textsc{Fornax}~\citep{Radice+2017,Vartanyan+2018,Burrows+2021,Vartanyan+2021,Vartanyan+2023a} obtained neutrino-driven explosions, which were often fostered by strong SASI shock sloshing along the symmetry axis of the computational grid, because SASI mass motions were amplified in a positive feedback process with enhanced neutrino heating in the polar directions defined by the grid axis, in particular with the RbR+ transport (as correctly pointed out in References~\citenum{Dolence+2015,Skinner+2016}). In contrast, the \textsc{Castro} code~\citep{Dolence+2015}, \textsc{Vulcan} code, and the Japanese Boltzmann-transport code~\citep{Nagakura+2019}, all using Newtonian gravity, did not yield neutrino-driven explosions. Instead, the \textsc{Vulcan} models suggested the possibility of a new acoustic mechanism, in which acoustic energy flux generated by large-amplitude PNS oscillations powered rather late and weak CCSN explosions \citep{Burrows+2006a,Burrows+2006b,Burrows+2007,Ott+2008}. Independent MD simulations, however, could not confirm the viability of this alternative mechanism, because the acoustic energy flux, either produced by PNS vibrations or PNS convection~\citep{Gossan+2020} or by accretion downflows impacting onto the PNS surface~\citep{Janka+2008}, was far too weak to make a significant effect on the shock evolution; the acoustic energy flux was not able to compete with the much stronger energy transfer to the shock by neutrino-energy deposition \citep{Janka+2008,Marek+2009,Wang+2024}.

The mentioned artifacts introduced by the symmetry axis of 2D SN models, possibly amplified in connection with RbR transport, are one of the reasons for pushing toward 3D simulations, where RbR transport artifacts are much less worrisome~\citep{Glas+2019a}. The toroidal structures of 2D models (see images in~\citenum{Couch2013}) are unnatural and prevent direct comparisons, quantitatively as well as qualitatively, with observed CCSN asymmetries and radial mixing of chemical elements~\citep{Hammer+2010}. Moreover, the power spectrum of fragmenting and cascading flow vortices and eddies in the quasi-turbulent neutrino-heated postshock layer exhibit fundamental differences in 2D and 3D (see detailed discussion in \citenum{Hanke+2012}). All of these points of concern demand the lifting of CCSN models from 2D to 3D.

\subsubsection{Supernova Models Arriving in the Third Dimension}

The first self-consistent, first-principle 3D simulations of neutrino-driven CCSN explosions with multigroup neutrino transport were based on these numerical developments by straightforwardly generalizing the codes from 2D to 3D and coping with the roughly 100 times higher computational demands in 3D by utilizing the first sufficiently powerful massively parallel supercomputers that became available. Since the previous review in this journal by Janka, Melson, \& Summa~\citep{Janka+2016}, 3D CCSN modeling with elaborate neutrino transport has developed from its infancy to a mature state, and the pool of 3D models has expanded accordingly.

Neutrino-driven explosions were thus obtained for progenitors of 11.2\,$M_\odot$~\citep{Takiwaki+2012,Mueller2015}, confirming the successful explosions of this progenitor model previously found in 2D, of 9.6\,$M_\odot$~\citep{Melson+2015a}, 15\,$M_\odot$~\citep{Lentz+2015}, 20\,$M_\odot$ (\citenum{Melson+2015b}; slightly modified neutral-current neutrino-nucleon scattering rates due to strangeness-dependent contributions to the axial-vector coupling constant were crucial for this successful explosion), 18\,$M_\odot$~\citep{Mueller+2017} and 18.88\,M$_\odot$~\citep{Bollig+2021}, which exploded only when 3D perturbations in the convective oxygen-burning shell of the progenitor were taken into account~\cite{Mueller+2016,Yadav+2020}, for a small selection of progenitors between 12\,$M_\odot$ and 40\,$M_\odot$~\citep{Ott+2018}, for a suite of ultra-stripped~\citep{Mueller+2018} and low-mass single-star and binary progenitors~\citep{Mueller+2019,Glas+2019a,Stockinger+2020,Wang+2024}, and for rapidly extended, large sets of progenitors over a wide mass range between about 9\,$M_\odot$ and 100\,$M_\odot$~\citep{Vartanyan+2019a,Burrows+2019,Burrows+2020,Vartanyan+2022,Burrows+2024,Burrows+2024a,Nakamura+2025}. 

Despite the general agreement that the neutrino-driven mechanism is able to drive CCSN explosions, simulations for individual progenitors often differ quantitatively as well as qualitatively (see also Section~\ref{sec:explodability}). This, however, is not astonishing in view of the different hydrodynamics codes and grids, different employed nuclear EoSs, different choices of gravity, and different methods for the neutrino transport and the descriptions of the neutrino processes. An attempt to compare existing 2D CC results for the same progenitors referring to information in the literature can be found in~\citep{OConnor+2018b}. Unfortunately, a community-involving code comparison in greater detail, similar to the 1D project of~\citep{OConnor+2018c} and the sporadic efforts by individual groups or collaborations \citep{Liebendoerfer+2005,Mueller+2010,Just+2018,Varma+2021}, using well controlled numerical setups and (systematically) varied inputs to assess the most critical aspects of neutrino transport and microphysics, does not yet exist for MD simulations.

\begin{figure}[htbp]
\center
\includegraphics[width=15.0cm]{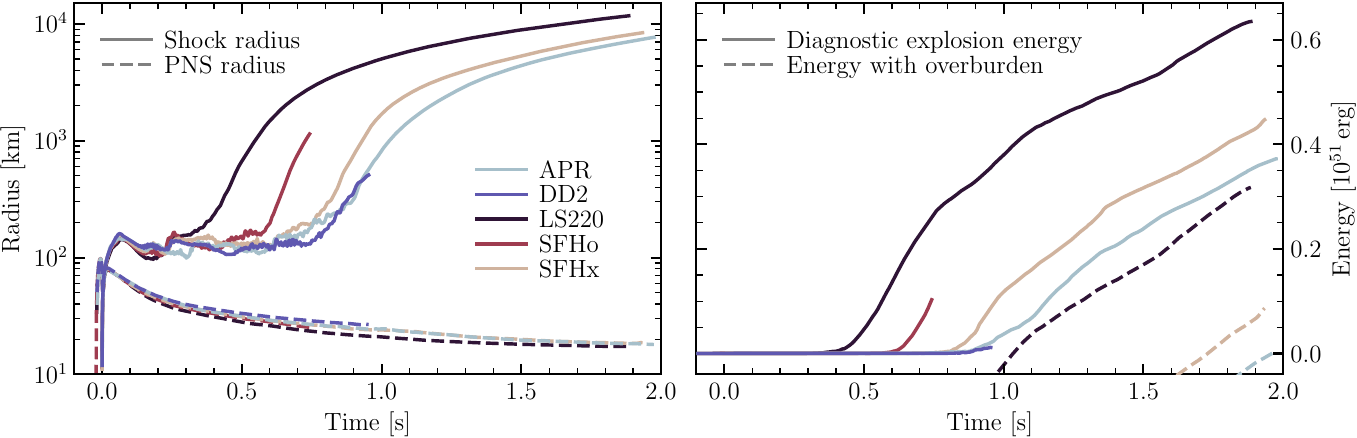}
\caption{Equation-of-state dependence of CCSN explosions of an 18.88\,$M_\odot$ star (R.~Bollig, private communication, and References~\citenum{Bollig+2021,Janka+2023}; Kresse et al., in preparation). All 3D SN simulations were started from a progenitor model with 3D perturbations of velocity, density, and chemical composition due to convective oxygen-shell burning~\citep{Yadav+2020} and employed different nuclear EoSs widely used in SN simulations: LS220~\citep{Lattimer+1991}, SFHo, SFHx~\citep{Steiner+2013,Hempel+2010}, DD2~\citep{Typel+2010,Hempel+2012}, and APR~\citep{Schneider+2019b}. {\em Left panel:} Spherically averaged shock radius (solid lines) and PNS radius (defined at a density of $10^{11}$\,g\,cm$^{-3}$; dashed lines) as functions of post-bounce time. {\em Right panel:} Time evolution of diagnostic explosion energy (solid lines) and explosion energy with overburden energy of the progenitor taken into account (dashed lines; for a definition of these energies, see Section~\ref{sec:lessons}). In all cases successful neutrino-driven explosions were obtained, but there is a clear correlation between the onset of the explosion and the contraction of the PNS. The faster the PNS contracts, the earlier the explosion sets in. This signals a crucial and sensitive influence of the PNS's radius evolution on the start of the explosion. Note that the PNS radii in the simulations with the APR and SFHx EoSs nearly overlap, and the shocks and explosion energies in both cases show similar behavior. Figure courtesy of Daniel Kresse and Robert Bollig.}
\label{fig:EoSdependence}
\end{figure}

\subsubsection{Lessons Learned}
\label{sec:lessons}

In summary, a variety of microphysics effects turned out to be supportive for explosions in MD simulations, for example  GR~gravity~\citep{OConnor+2018b} and many-body corrections (i.e., in-medium effects) in neutrino-nucleon interactions~\citep{Horowitz+2017,Burrows+2018}, both of which had been recognized to be of great relevance in previous 1D and 2D studies already (e.g., \citenum{Bruenn+2001,Liebendoerfer+2005,Buras+2006,Buras+2006b}). The deeper gravitational potential in GR leads to a more compact and hotter PNS, which radiates higher neutrino luminosities and harder neutrino spectra, thus causing stronger postshock neutrino heating. Many-body corrections effectively reduce the neutral-current neutrino-nucleon scattering opacity and thus allow the heavy-lepton neutrinos to escape more easily from the PNS interior, which again triggers faster PNS contraction and this, in turn, enhances the $\nu_e$ and $\bar\nu_e$ emission and thus the postshock heating, fostering earlier explosions. There are numerous other possibilities on the microphysics side to obtain similar causal relationships in the SN core, for example by including a reduced neutrino-nucleon scattering cross section due to strangeness-dependent contributions to the axial-vector coupling~\citep{Melson+2015b}, the formation of muons in the hot PNS medium~\citep{Bollig+2017}, and a ``soft'' hot nuclear EoS (\citenum{Janka2012,Suwa+2013}; ``soft'' in the sense of leading to faster PNS contraction), e.g., facilitated by a large effective nucleon mass in the regime above nuclear saturation density to reduce the nucleonic thermal contributions to the gas pressure~\citep{daSilvaSchneider+2019,Yasin+2020}, or by a combination of incompressibility and symmetry-energy slope~\citep{Couch2013a}. 

Figure~\ref{fig:EoSdependence} demonstrates the crucial influence of the PNS contraction on the evolution of the SN shock by means of results from a set of recent 3D CCSN simulations with \textsc{Prometheus-Vertex}, employing different nuclear EoSs~\citep{Janka+2023}. The left panel shows the time evolution of the angle-averaged shock radii and the right panel the explosion energies. The so-called diagnostic explosion energy is defined as the integrated total energy of all matter in the gain layer behind the SN shock with positive total energy, i.e., positive values of internal plus kinetic plus gravitational energy. Here the internal energy includes thermal and degeneracy energies of the stellar plasma (without particle rest-mass energies) and the gravitational energy is evaluated with the Newtonian gravitational potential of the mass enclosed at each radius reduced by the mass-equivalent of the energy carried away by neutrinos until a considered time. This diagnostic energy is distinguished from the explosion energy that takes into account the overburden energy of the progenitor, i.e., the negative binding energy of all stellar matter ahead of the SN shock. Both energies converge at late times and asymptote to the potentially measurable energy of the SN explosion.

\begin{figure}[htbp]
\center
\includegraphics[width=15cm]{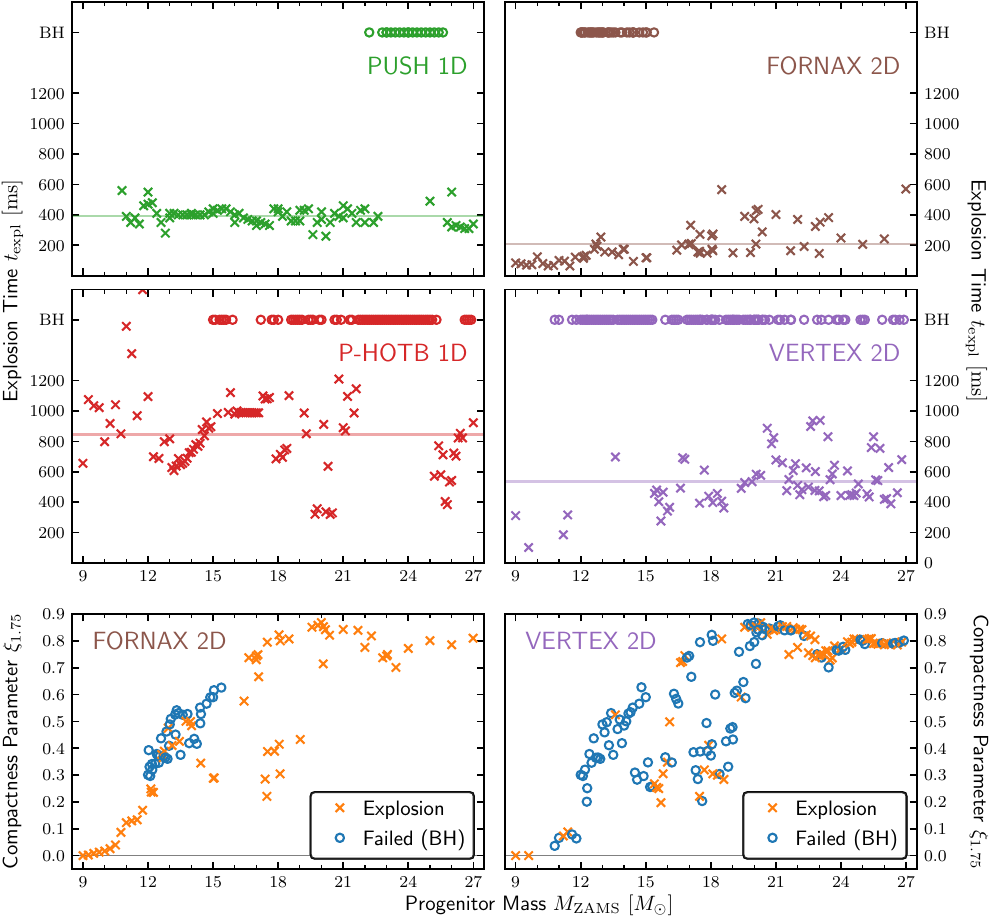}
\caption{``Explodability'' as function of the ZAMS mass of progenitor stars in large sets of CC simulations using different ``neutrino engines'' in 1D calculations and different neutrino-hydrodynamics codes in 2D models to obtain neutrino-driven explosions (Glas et al., in preparation). Circles mark BH forming cases and crosses indicate successful explosions. {\em Upper four panels:} Onset time of the explosion (defined by the post-bounce time when the average shock radius exceeds 300\,km) versus ZAMS mass with averages indicated by horizontal lines. The results of the \textsc{Push} 1D engine models ({\em top left}) were taken from~\citep{Ebinger+2019}, where progenitors from~\citep{Woosley+2002,WoosleyHeger2007} were considered; the \textsc{P-Hotb} 1D engine models ({\em middle left}) are from~\citep{Sukhbold+2016} using progenitors from~\citep{Sukhbold+2014,WoosleyHeger2015}; and the \textsc{Fornax} 2D simulations \citep[from][{\em top right}]{VartanyanBurrows2023} as well as the \textsc{Vertex} 2D simulations (R.~Bollig, private communication; {\em middle right}) employed the SFHo EoS~\citep{Steiner+2013,Hempel+2010} and progenitors for $\ge$12\,$M_\odot$ from~\citep{Sukhbold+2018} and for $<$12\,$M_\odot$ from~\citep{WoosleyHeger2015} and, in the case of \textsc{Vertex}, also from~\citep{Woosley+2002}. Both \textsc{Push} and \textsc{P-Hotb} triggered explosions artificially in 1D with neutrino engines whose parameters were calibrated by reproducing explosion properties of SN~1987A. \textsc{Push} yields 79 (81\%) explosions and 18 (19\%) BH cases out of 97 models, \textsc{P-Hotb} 90 (57\%) explosions and 67 (43\%) BHs in 157 models, \textsc{Fornax} 63 (63\%) explosions and 37 BH cases (37\%) of 100 models, and \textsc{Prometheus-Vertex} 73 (41\%) explosions and 104 BH cases (59\%) of 177 models. {\em Bottom panels:} Core compactness $\xi_{1.75}$ versus ZAMS mass for the \textsc{Fornax} and \textsc{Vertex} 2D simulations with successful and failed explosions indicated by different symbols. Gray horizontal lines mark $\xi_{1.75} = 0$ to guide the eye. Figure courtesy of Robert Glas.}
\label{fig:explodability}
\end{figure}

\subsection{Stellar Explodability}
\label{sec:explodability}

Although the neutrino-heating mechanism is widely accepted as driver of most CCSN explosions now, its success or failure in individual progenitor stars (progenitor ``explodability'') is still controversial, and therefore the progenitor-explosion connection expressed by the ``landscape'' of NS and BH forming events depending on the progenitor properties is still uncertain. MD simulations with different codes yield different answers because of the complexity of the involved physics and because the explosion as a threshold phenomenon can be very sensitive to smaller details. Similarly, 1D ``neutrino engines'', which are designed to artificially initiate explosions, yield different answers because of different methods for the explosion trigger and considerable sensitivity to the choice of associated parameter values. For example, such engines use enhanced neutrino heating regulated by an efficiency parameter~\citep{OConnor+2011}; or they tap energy from the heavy-lepton neutrino fluxes~\citep[\textsc{Push};][]{Ebinger+2019}, tune or calibrate the $\nu_e$ and $\bar\nu_e$ luminosities from the PNS core~\citep[\textsc{P-Hotb};][]{Sukhbold+2016}, model CCSNe semi-analytically including parametrized MD effects~\citep{Mueller+2016a}, exploit energy input from a PNS's neutrino-driven wind (NDW; \citenum{Pejcha+2015}), or employ a mixing-length treatment (MLT) of postshock turbulence (\textsc{Stir}; \citenum{Couch+2020}).
\begin{marginnote}[]
\entry{NDW}{Neutrino-driven winds (``neutrino winds'') are outflows of baryonic matter blown off the surface of a hot PNS by neutrino heating}
\end{marginnote}
\begin{marginnote}[]
\entry{MLT}{Mixing-length theory of convection}
\end{marginnote}

Figure~\ref{fig:explodability} displays published results from two such 1D neutrino engines, \textsc{Push} and \textsc{P-Hotb}, for older and newer progenitor sets. Obviously, there are significant differences concerning success or failure of explosions as functions of the zero-age-main-sequence (ZAMS) mass, and in the case of neutrino-driven shock revival also significant differences exist in the explosion time, implying different predictions for the NS masses. Most of these differences are a consequence of the different engine treatments, but some may also be connected to the inconsistent progenitor sets. Progenitor models from stellar evolution calculations by different groups actually lead to largely different predictions for NS and BH forming CC events, as was shown in~\citep{Boccioli+2023} for the same engine prescription (\textsc{Stir}).
\begin{marginnote}[]
\entry{ZAMS mass}{Zero-age-main-sequence mass meaning the birth mass of stars on the main sequence in the Hertzsprung-Russell diagram}
\end{marginnote}

\begin{marginnote}[]
\entry{Compactness}{The compactness of a stellar core of mass $M$ enclosed by radius $R$ is defined as $\xi_M = (M/M_\odot)/(R(M)/1000\,\mathrm{km})$ \citep{OConnor+2011}. Here the standard choice is $M = 1.75\,M_\odot$}
\end{marginnote}
Figure~\ref{fig:explodability} also presents results from full (180$^\circ$) 2D simulations with two state-of-the-art CCSN codes, \textsc{Fornax} and \textsc{Prometheus-Vertex}, thus comparing two different codes applied to progenitors drawn from the same set of stellar evolution models. The outcomes with ZAMS mass and core compactness $\xi_{1.75}$ are considerably different in both cases. \textsc{Fornax} yields significantly earlier explosions for a much larger fraction of considered progenitors than \textsc{Prometheus-Vertex}, despite the fact that the use of the RbR+ neutrino transport in \textsc{Vertex} tends to facilitate axis-dominated explosions in 2D. Considering the exactly same progenitors, the results with \textsc{Prometheus-Vertex} are consistent with simulations with the \textsc{Alcar} code when the RbR+ transport version in this code is used (\citenum{Just+2018}; Glas et al., in preparation), whereas \textsc{Fornax} produces explosions in many cases where \textsc{Prometheus-Vertex} and \textsc{Alcar} do not find them. 

It is particularly interesting that \textsc{Fornax}, much different from \textsc{Prometheus-Vertex}, obtains failed explosions only in a mass window between roughly 12\,$M_\odot$ and 15.5\,$M_\odot$ and yields explosions for all progenitors above and below this mass interval, in particular also for all progenitors with a high compactness of the stellar core ($\xi_{1.75}$ in Figure~\ref{fig:explodability}). This was not expected on grounds of the 1D neutrino-engine modeling except with the \textsc{Stir} method~\citep{Boccioli+2023}.  

The detailed reasons for the different 2D results with the two neutrino-hydrodynamics codes are not identified. Nucleon correlations~\citep{Burrows+2018,Horowitz+2017}, insufficient grid resolution, and constrained dimensionality~\citep{Nordhaus+2010a} have been suggested but can be excluded as explanations. Many-body effects in the neutrino-nucleon interactions are also taken into account in \textsc{Prometheus-Vertex} (for both neutral-current and charged-current reactions and including density and temperature dependence according to~\citenum{Burrows+1998,Burrows+1999}, see~\citenum{Buras+2006}) and \textsc{Alcar}; resolution changes have been tested to not affect the conclusions; and 3D is unlikely to provide better conditions for explosions, because \textsc{Prometheus-Vertex} yields explosions more readily in 2D (in agreement with results in References \citenum{Hanke+2012,Takiwaki+2012,Takiwaki+2014,Couch+2013}). In contrast, 3D effects were found to strongly support explosions in \citep{Nordhaus+2010} and less strong effects in the same direction were witnessed in \citep{Dolence+2013,Mueller2015}, whereas no differences in the explosion behavior between 2D and 3D were reported for \textsc{Fornax}~\citep{Burrows+2021}. The lack of 2D-3D differences in the explodability is another indicator, besides the earlier explosion times, that \textsc{Fornax} produces significantly stronger neutrino-driven explosions than \textsc{Prometheus-Vertex}. 

In fact, the reduced tendency to get explosions with \textsc{Prometheus-Vertex} in 3D has an important consequence. Unless one considers the lowest-mass CCSN progenitors with ZAMS masses around (9--$10)\,M_\odot$~\citep{Melson+2015a,Stockinger+2020} or includes rapid rotation~\citep{Summa+2018} or slightly modified weak neutral-current interactions (\citenum{Melson+2015b}; see also Section~\ref{sec:MDhistory}), \textsc{Prometheus-Vertex} yields neutrino-driven explosions in 3D only when the simulations are started from 3D progenitor conditions, i.e., when the final stages (minutes to hours) of the convective oxygen-shell burning have been computed in 3D~\citep{Mueller+2016,Yadav+2020} and the corresponding density and velocity perturbations in the O-shell have sufficiently large amplitudes in sufficiently extended radial shells~\citep{Mueller+2017,Bollig+2021,Janka+2024}. Large-scale pre-collapse velocity perturbations can lead to density variations up to more than 20\% and non-radial velocities of several 1000\,km\,s$^{-1}$ when the O-shell falls through the shock. Since the stalled shock expands more easily into cavities of lower density and lower ram pressure, this causes time-dependent shock deformation. Oblique mass flow through the shock and considerable non-radial velocities of the accreted matter act as boosters of postshock convection and turbulence~\citep{Mueller+2015,Mueller+2017,Mueller2020,Vartanyan+2022}. Magnetic fields, if they are sufficiently strong in the pre-collapse stellar core or grow sufficiently rapidly in the turbulent postshock flow, can also assist the onset of the explosion even in nonrotating stars, among other effects due to additional magnetic energy and pressure~\citep{Obergaulinger+2014,Mueller+2020,Varma+2023,Sykes+2024,Nakamura+2025}. CC simulations with the \textsc{Fornax} code produce explosions for a wide variety of progenitors in 2D and 3D without any such additional support~\citep{Burrows+2020,  Burrows+2021,VartanyanBurrows2023,Burrows+2024}.


\begin{figure}[h]
\center
\includegraphics[width=15cm]{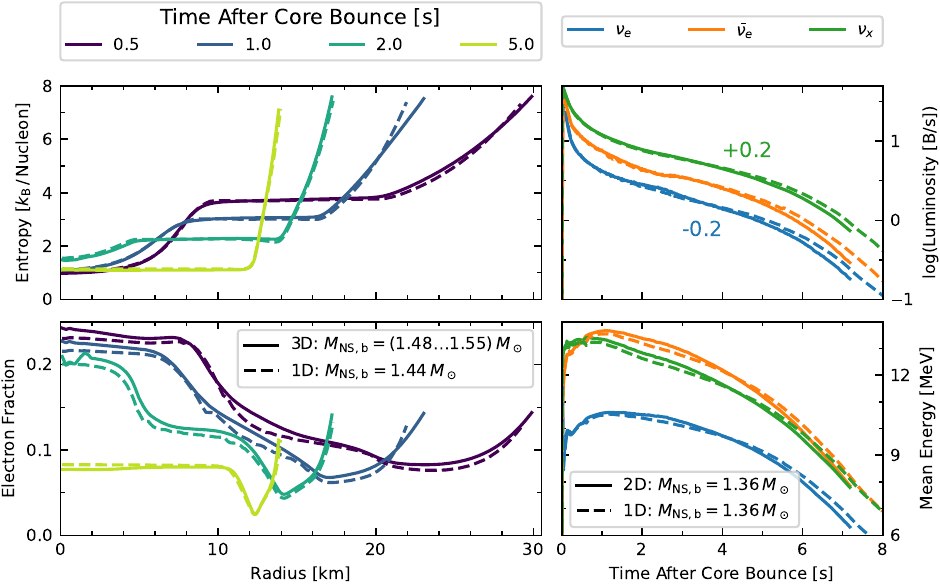}
\vspace{-5pt}
\caption{PNS convection in MD SN simulations compared to MLT convection in 1D PNS cooling calculations with the \textsc{Prometheus-Vertex} code (Heinlein et al., in preparation). {\em Left panels:} Radial profiles of matter entropy and electron fraction at different post-bounce times (denoted in the box above the panels) for a 1D PNS model with a baryonic mass of 1.44\,$M_\odot$ and a 3D simulation of a PNS with a baryonic mass growing from 1.48\,$M_\odot$ at 0.5\,s to 1.55\,$M_\odot$ at 5.0\,s. The PNS is defined by a mass-density $\rho \ge 10^{11}$\,g\,cm$^{-3}$ and contracts from initially 30\,km to 14\,km. Small differences between the 1D and the 3D model near the surface are a consequence of continuous accretion in the 3D case. The convective layer can be recognized by flat entropy profiles. {\em Right panels:} Lab-frame luminosities ({\em top}) and mean energies ({\em bottom}) of radiated $\nu_e$, $\bar\nu_e$, and a single species of heavy-lepton neutrinos, $\nu_x$, as functions of time for a 1D PNS cooling model with MLT convection compared to a 2D simulation that was chosen because its constant baryonic PNS mass of 1.36\,$M_\odot$  closely matches that of the 1D model. Note that the luminosities of $\nu_e$ and $\nu_x$ are shifted by constant factors to avoid cluttering of the different lines. The 2D neutrino data are angle-averaged. MLT convection is able to reproduce the results obtained by MD simulations with good accuarcy; here it is important to know that there are no appreciable differences between PNS convection in 2D and 3D. Figure courtesy of Malte Heinlein.}
\label{fig:PNSconvection}
\end{figure}

\section{LONG-TERM SUPERNOVA MODELING}
\label{sec:longtermSNe}

Most of the CCSN codes mentioned in Section~\ref{sec:MDhistory} have been applied so far only to relatively short-time computations, covering periods of just tens of milliseconds after core bounce in 3D for the most expensive methods with Boltzmann transport and a few hundred milliseconds up to around one second in the majority of 2D and 3D applications of RbR and MD AEF two-moment schemes. Long-term simulations that follow the evolution with detailed neutrino treatment for many seconds after bounce are the new frontier but still pose a grand computational challenge by requiring huge amounts of supercomputing resources. Meanwhile there are several undertakings to extend MD neutrino-hydrodynamics simulations over periods of many seconds, first with a 2D-1D hybrid approach~\citep{Suwa2014}, full 2D~\citep{Nakamura+2019,Burrows+2021,Bruenn+2023}, and full 3D~\citep{Mueller+2019,Bollig+2021,Burrows+2024}. These simulations obtained self-consistent neutrino-driven explosions that begin to serve as starting points for purely hydrodynamic extension simulations that follow the shock evolution until shock-breakout from the stellar surface~\citep{Mueller+2018,Chan+2020,Stockinger+2020,Rahman+2022,Sandoval+2021,Vartanyan+2025,EggenbergerAndersen+2024,Sykes+2024a}, thus upgrading previous similar 2D and 3D works that initiated neutrino-driven explosions with tuned neutrino luminosities in a more approximate description of neutrino transport and postshock heating \citep{Kifonidis+2003,Hammer+2010,Wongwathanarat+2013,Wongwathanarat+2015,Wongwathanarat+2017,Gabler+2021}.

With the currently most rigorous treatments of neutrino transport (AEF and RbR+ two-moment schemes with full energy-bin coupling and all velocity-dependent terms), such computational models require $\gtrsim$10$^7$ core hours for $\sim$1\,s of evolution with medium resolution (equivalent to angular bins of $\sim$2$^\circ$) and several times more with high resolution. All current attempts employ ways to reduce or cap the needed resources by either saving on the numerical resolution, on the detailedness of the neutrino transport, or on the microphysics taken into account. These differences should be carefully considered and kept in mind when results are compared and conclusions from those are drawn. In the following, the modeling strategy of the long-time simulations with \textsc{Prometheus-Vertex} will be briefly recapitulated.

\subsection{The \textsc{Prometheus-Vertex} and \textsc{Prometheus-Nemesis} Codes}
\label{sec:PVNmethods}

The \textsc{Prometheus-Vertex} code comprises the higher-order Godunov-type \textsc{Prometheus} hydrodynamics module with an exact Riemann solver~\citep{Fryxell+1991,Mueller+1991} combined with the \textsc{Vertex} neutrino transport code~\cite{Rampp+2002}. \textsc{Vertex} solves the multigroup velocity-dependent neutrino transport in the comoving frame of the stellar plasma. It can handle three ($\nu_e$, $\bar\nu_e$, and $\nu_x$ representing heavy-lepton neutrinos), four (discriminating $\nu_x$ and $\bar\nu_x$) or six (additionally discriminating muon and tau neutrinos when muons are included in the medium) neutrino species and applies the RbR+ approximation in 2D and 3D simulations. It solves the 1D transport (dependent on fully coupled energy groups and the propagation angle relative to the radius vector) in each angular direction of the spatial grid through the convergent iteration of the two-moment equations and the Boltzmann equation. Although this approximation assumes the neutrino phase-space distribution to be axially symmetric around the radial direction (i.e., only the propagation angle relative to the radial direction is considered), which implies that only radial fluxes exist, non-radial advection of trapped neutrinos moving with the stellar plasma and the non-radial components of the neutrino pressure gradients are taken into account (indicated by the ``+'' suffix to RbR).

\textsc{Vertex} is supplemented with a state-of-the-art treatment of the energy-dependent neutrino reaction rates with matter particles (nuclei, nucleons, electrons, positrons, and optionally muons and anti-muons) as well as neutrino-neutrino interactions and neutrino-antineutrino pair creation and annihilation by nucleon-nucleon bremsstrahlung (summaries reporting successive upgrades are given in \citenum{Rampp+2002,Buras+2003,Langanke+2003,Buras+2006,Janka+2007,Langanke+2008,Janka2012,Bollig+2017,Fiorillo+2023}). The charged-current as well as neutral-current neutrino-nucleon interactions include many-body corrections due to nucleon correlations, weak magnetism, and recoil, and the charged-current processes also account for the energy shifts due to potential-energy differences between neutrons and protons in the medium. Neutrino-lepton interactions (also those between neutrinos) include scattering processes as well as pair annihilation.

The \textsc{Nemesis} (Neutrino-Extrapolation Method for Efficient SImulations of Supernova explosions) neutrino treatment, which was first introduced in~\citep{Stockinger+2020} and further refined in \citep{Kresse2023}, was developed to replace \textsc{Vertex} in order to reduce the huge requirements of computational resources for long-term 3D CCSN simulations (in practise, the gain is a factor of 10--30 with the greater saving when the simple nuclear flashing treatment of Reference~\citenum{Rampp+2002} is used in the non-NSE regime instead of a nuclear network). \textsc{Nemesis} is able to essentially seamlessly continue the simulations conducted with \textsc{Vertex} without any disturbing numerical transients. It employs results from 1D PNS cooling simulations with \textsc{Prometheus-Vertex}, which include a MLT treatment for the lepton-number and energy fluxes due to convection in the PNS interior (see equations~3 and~4 in Reference~\citenum{Mirizzi+2016}, corrected on the right-hand sides by factors $-m_B^{-1}$ and $-1$, respectively, with $m_B$ being the baryonic unit mass). MLT convection in 1D PNS cooling simulations yields excellent agreement both of PNS and of neutrino properties compared to MD simulations (Figure~\ref{fig:PNSconvection}).

\begin{figure}[t]
\center
\includegraphics[width=16cm]{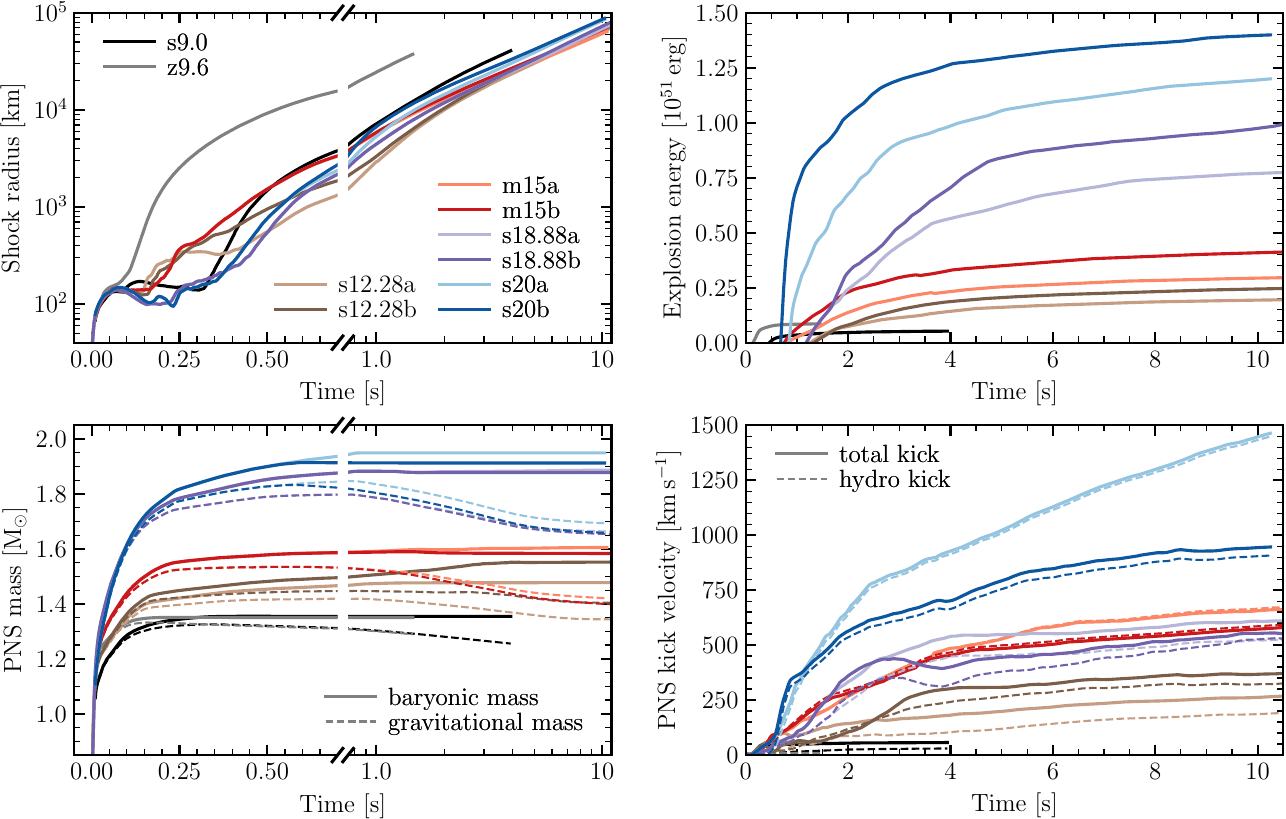}
\caption{Long-term 3D CCSN simulations of the Garching group for 9.0, 9.6, 12.28, 15, 18.88, and 20\,$M_\odot$ progenitors (\citenum{Stockinger+2020,Bollig+2021,Janka+2024}; Kresse et al., in preparation; models~s12.28b, m15a, m15b, s18.88b, s20a, and s20b correspond to models s12.28, m15, m15e, s18.88, s20, and s20e, respectively, already discussed in Reference~\citenum{Janka+2024}). The calculations were performed with the \textsc{Prometheus} hydrodynamics code including \textsc{Vertex} neutrino transport for periods between 0.5\,s and 5.1\,s and continued with the \textsc{Nemesis} neutrino treatment to later times. The m15 models are based on a rotating progenitor~\citep{Summa+2018}, and the s12.28 and s18.88 models made use of progenitors whose convective oxygen-shell burning had been computed in 3D for the final hour before CC in the 12.28\,$M_\odot$ case and the final 10\,min in the 18.88\,$M_\odot$ case~\citep{Yadav+2020}. The CCSN runs were either performed with the LS220 EoS~\citep{Lattimer+1991} or the SFHo EoS~\citep{Steiner+2013,Hempel+2010} \citep[for details, see][]{Janka+2024}. {\em Top left:} Spherically averaged shock radii as functions of post-bounce time. {\em Bottom left:} PNS baryonic masses (solid lines) and gravitational masses (dashed lines) as functions of post-bounce time. {\em Top right:} Explosion energies versus post-bounce time with overburden energies (i.e., the binding energies of the progenitor stars above the outward moving shocks) taken into account (for a definition, see Section~\ref{sec:lessons}). The a and b versions of the models differ slightly in their explosion energies (b versions being more energetic) and roughly bracket uncertainties connected to neutrino heating and cooling as well as hydrodynamic stochasticity. {\em Bottom right:} Time evolution of the PNS kicks due to asymmetric mass ejection (dashed lines) and asymmetric mass ejection plus anisotropic neutrino radiation (solid lines). Figure courtesy of Daniel Kresse.}
\label{fig:3Dmodels}
\end{figure}

\textsc{Nemesis} obviates the time-consuming neutrino transport in a 3D calculation by referring to the $\nu_e$ and $\bar\nu_e$ luminosities and mean energies from a 1D PNS evolution model of similar mass and using them in the neutrino heating rates as well as the lepton-number transfer rates from neutrinos to matter outside the PNS. These rates and their inverse reactions are properly scaled with the neutrino quantities and the local medium conditions (temperature, density, electron fraction) in the heating and cooling layers exterior to the neutrinospheres. The employed formulation of the neutrino luminosities includes an accretion contribution depending on the PNS mass accretion rate, which is an effect missing in the 1D PNS models. In the interior of the PNS, which is treated in 1D with MLT convection in the \textsc{Nemesis} applications, the time-dependent cooling profiles from the 1D PNS cooling model are adopted and neutrino pressure is taken into account. This ensures that the PNS contraction follows closely a simulation with detailed neutrino transport, and the time-dependent PNS and gain radii of full 3D simulations are well reproduced. Tests also confirmed that the ejecta and explosion properties of 3D simulations with \textsc{Nemesis} show good agreement with 3D results obtained with \textsc{Vertex} neutrino transport. Important aspects of the 3D SN evolution can thus be treated more efficiently and with considerably reduced computational expenses using the \textsc{Nemesis} approximation of neutrino effects.

\begin{figure}[!]
\center
\includegraphics[width=16cm]{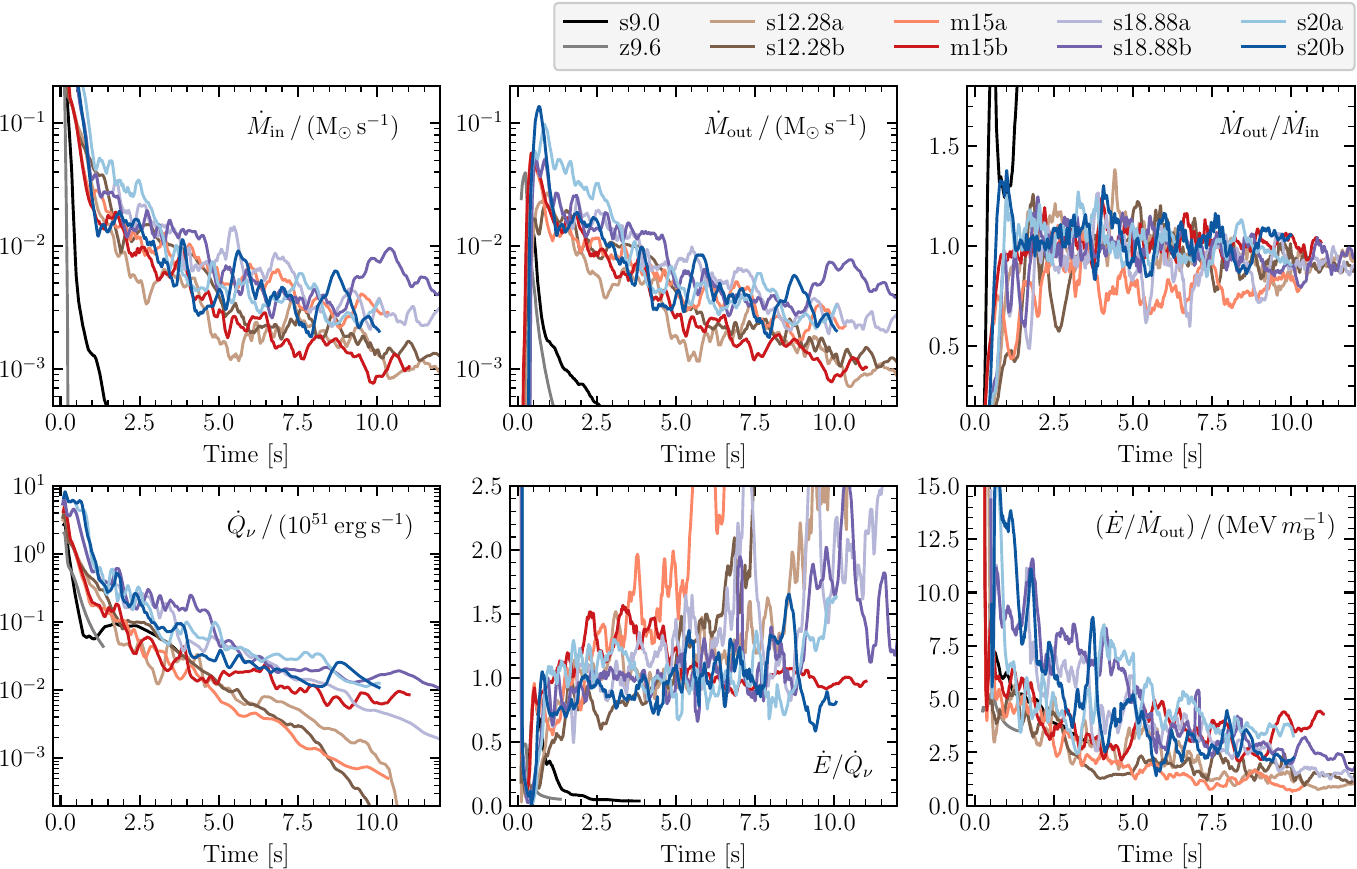}
\caption{Mass downflow and outflow properties, neutrino heating rates, and explosion energetics in the long-term 3D CCSN simulations of 9.0, 9.6, 12.28, 15, 18.88, and 20\,$M_\odot$ progenitors shown in Figure~\ref{fig:3Dmodels} (\citenum{Bollig+2021,Kresse2023}; Kresse et al., in preparation). {\em Upper row:} Mass downflow rates ({\em left}), mass outflow rates ({\em middle}), and ratios of outflow rate to inflow rate as functions of post-bounce time ({\em right}), all measured at a radius of 400\,km. Inflow and outflow rates are effectively equal between $\sim$1\,s and $\sim$(6--8)\,s, because inflowing matter absorbs energy from neutrinos close to the PNS and is re-ejected afterwards as long as neutrino heating is still strong enough. Afterwards $\dot M_\mathrm{out}/\dot M_\mathrm{in}$ begins to decrease, thus signaling the onset of fallback accretion by the PNS. Models s9.0 and z9.6 are exceptions as their mass inflows to the PNS stop early, giving way to NDW outflows, which, however, are much weaker than the inflow-outflow rates of all other models. Model m15a is also an exception with an outflow-to-downflow ratio smaller than unity, signaling continuous mass accretion by the PNS (see lower left panel of Figure~\ref{fig:3Dmodels}). {\em Lower row:} Total (i.e., volume-integrated) neutrino-heating rates ({\em left}), ratios of the growth rate of the explosion energy (with overburden taken into account) to the total neutrino-heating rate ({\em middle}), and ratios of the growth rate of the explosion energy ($\dot E$ with overburden taken into account) to the mass outflow rate at 400\,km ({\em right}). The energy deposited by neutrino heating around the PNS leads to a continuous, long-lasting growth of the explosion energy (see upper right panel of Figure~\ref{fig:3Dmodels}). Models s9.0 and z9.6 are again exceptions, because their neutrino heating achieves to gravitationally unbind near-surface PNS matter in the NDWs, however without providing any relevant net gain for the explosion energies. In phases where the growth rate of the explosion energy exceeds the total rate of neutrino energy deposition, $\dot E/\dot Q_\nu > 1$, nuclear energy generation contributes to the growth of the explosion energy. This typically happens at very late times ($\gtrsim$6\,s after bounce) except in models~m15a and m15b, where the heating rates drop more rapidly in comparison. Figure courtesy of Daniel Kresse.}
\label{fig:3Dmodels-outflows}
\end{figure}

\subsection{New Insights From Long-Term Supernova Simulations}
\label{sec:longterm-insights}

\subsubsection{Explosion Energies}
\label{sec:expergs}

Long-term simulations are indispensable to determine the observable energies of CCSNe. Figure~\ref{fig:3Dmodels} displays the observationally relevant explosion energies for a set of 3D models computed with \textsc{Prometheus-Vertex/Nemesis}. As stated in Section~\ref{sec:explodability}, the self-consistent successful explosions in 3D were obtained under special assumptions (low-mass progenitors, rotation, slightly modified neutrino opacities) or when 3D progenitor conditions were considered  \citep[a detailed description can be found in][]{Janka+2024}. The plotted energies as functions of time, $E(t)$, include the binding energy of the stellar layers exterior to the outward traveling SN shock. This negative energy of the outer shells of the progenitor star is added to the ``diagnostic'' explosion energy, which is often provided for short-time simulations and comprises only the integrated energy of all ejected postshock matter with positive total (i.e., internal plus kinetic plus gravitational) specific energy $e_\mathrm{tot}$ (observable and diagnostic energies in comparison can be seen in Figure~\ref{fig:EoSdependence}). 

In explosions of progenitors with oxygen-neon cores (so-called electron-capture SNe) or low-mass iron cores with similarly steep outer density gradients (``ECSN-like'' events, see Reference~\citenum{Mueller2016}) the energies effectively saturate within the first second after core bounce. After the initiation of the neutrino-driven explosion, assisted by a short phase ($\sim$0.1\,s) of post-bounce convection~\citep{Melson+2015a,Stockinger+2020,Janka+2023,Wang+2024}, neutrino heating near the PNS surface produces an essentially spherical baryonic wind~\citep{Qian+1996}. This NDW adds a minor contribution to the final explosion energy, because the wind mass-loss rates are low and a major fraction of the deposited neutrino energy is consumed to lift the wind material out of the deep gravitational potential of the PNS.
\begin{marginnote}[]
\entry{ECSN}{Electron-capture SN of a progenitor with a degenerate oxygen-neon core, whose gravitational collapse is triggered by electron captures and whose explosion is facilitated by the steep outer density gradient of the core}
\end{marginnote}

In contrast, in explosions of higher-mass (iron-core) progenitors long-lasting, massive accretion downflows exist and channel infalling gas close to the PNS, where this matter absorbs energy from the PNS's intense neutrino radiation and gets ejected again in outflows. Because of the high mass-inflow and outflow rates (see Figure~\ref{fig:3Dmodels-outflows}) and the gravitationally only weakly bound initial state of the infalling gas, the explosion energies can strongly rise over periods of several seconds, followed by a more gradual further growth that can continue for more than 10\,s. The explosion energy increases according to $\dot E = \dot M_\mathrm{out}\overline{h}_\mathrm{out}$, where $\overline{h}_\mathrm{out} = \overline{e}_\mathrm{tot} + \overline{P/\rho}$ is the specific ``total enthalpy'' ($P$, $\rho$ are pressure and mass-density, respectively), averaged over the mass outflow with rate $\dot M_\mathrm{out}$ (all measured at 400\,km). When the explosion begins, the integrated neutrino-energy deposition rate in the gain region, $\dot Q_\nu$, exceeds the growth rate of the explosion energy ($\dot E/\dot Q_\nu < 1$), because the energy transferred by neutrinos is mostly used to gravitationally unbind the matter initially sitting in the gain layer (but nuclear recombination energy of this material also adds to $\dot E$). Then a longer period of time follows where $\dot E \approx \dot Q_\nu$, consistent with neutrinos providing the bulk of the explosion energy during the phase of its steepest rise and largest growth. During this phase the downflows contain stellar matter falling inward from large initial radii where the gravitational binding energy is low. Finally and with a model-dependent appearance when $\dot Q_\nu$ has dropped to low values, the ratio becomes $\dot E/\dot Q_\nu > 1$. This signals that energy produced by nuclear reactions in the ejecta contributes significantly at late times, typically on the level of about (1--1.5)\,MeV per nucleon (Figure~\ref{fig:3Dmodels-outflows}, lower right panel), corresponding to the conversion of low-mass elements (C, O, Ne) in the downflows to more stable intermediate-mass (e.g., Si, S, Ar, C) and iron-group elements in the outflows. 

A larger set (20 models in total) of recent long-term 3D simulations of neutrino-driven explosions~\citep{Burrows+2024} demonstrates correlations, partly with considerable scatter, of explosion energy and NS (gravitational) mass with post-bounce explosion time, ejecta mass and energy dipoles, compactness $\xi_{1.75}$, total NS kick, and ejected mass of radioactive $^{56}$Ni, and with other quantities that are characteristic of the pre-collapse stellar structure and relevant for SN nucleosynthesis. Some of these interesting results confirm trends that had been expected on grounds of previous 2D CCSN calculations~\citep{Nakamura+2015}, parametric 1D explosion models \citep{Sukhbold+2016,Mueller+2016a,Ertl+2020,Pejcha2020}, and observations~\citep{MuellerT+2017}.

\subsubsection{Neutron Star and Black Hole Kicks and Spins}
\label{sec:kicks}

Asymmetric mass ejection in the explosion as well as asymmetric neutrino emission carry away linear momentum. Therefore momentum conservation implies that the new-born compact remnants in CCSNe can obtain natal kicks~\cite{Janka+1994,Burrows+1996}; in the presence of these two effects, asymmetric emission of GWs contributes only on a negligible level to these kicks because of the small mass-equivalent of the radiated GW energy ($E_\mathrm{GW}\lesssim 10^{-7}\,M_\odot c^2$ compared to $E_\nu^\mathrm{tot} \gtrsim 2\times 10^{53}\,\mathrm{erg}\sim 0.1\,M_\odot c^2$). A simple estimate of the hydrodynamic kick of a NS with (baryonic) mass $M_\mathrm{NS}$ yields~\cite{Janka2017b}
\begin{equation}
v_\mathrm{NS} \approx 211\,\mathrm{km\,s}^{-1}\,
\left(\frac{f_\mathrm{kin}}{\epsilon_5\,\beta_\nu}\right)^{\! 1/2}
\left(\frac{\alpha_\mathrm{ej}}{0.1}\right)\,
\left(\frac{E_\mathrm{exp}}{10^{51}\,\mathrm{erg}}\right)
\left(\frac{M_\mathrm{NS}}{1.5\,M_\odot}\right)^{-1} ,
\label{eq:vns3}
\end{equation}
where $\alpha_\mathrm{ej}$ is the momentum-asymmetry parameter of the SN ejecta and $E_\mathrm{exp}$ the explosion energy. The first factor is of order unity~\cite{Janka2017b}, and Eq.~(\ref{eq:vns3}) with this choice reproduces results of long-term CCSN simulations within some 10\%~\cite{Nakamura+2019}. Similarly, one estimates the neutrino-induced kick to be~\cite{Gessner+2018}
\begin{equation}
v_\mathrm{NS}^\nu \approx 167\,\mathrm{km\,s}^{-1}\,\,\frac{\bar{\alpha}_{\nu}^\mathrm{tot}}{0.005}\,\frac{E_\nu^\mathrm{tot}}{3\times 10^{53}\,\mathrm{erg}}\left(\frac{M_\mathrm{NS}}{1.5\,\mathrm{M}_\odot}\right)^{\!\! -1} \!\!,
\label{eq:vnuestimate}
\end{equation}
which means that a time-averaged asymmetry of the total neutrino energy loss ($E_\nu^\mathrm{tot}$ for $\nu_e$, $\bar\nu_e$, and all $\nu_x$) of 0.5\%, implying a dipole amplitude of 1.5\% of the monopole, leads to a recoil velocity of the NS of about 170\,km\,s$^{-1}$. The asymmetry of the neutrino loss expressed by the parameter $\bar{\alpha}_{\nu}^\mathrm{tot}$ emerges from anisotropic transport of neutrinos out of the neutron star interior, mainly connected to the LESA emission dipole (Lepton Emission Self-sustained Asymmetry; \citenum{Tamborra+2014,Janka+2016,OConnor+2018,Glas+2019,Vartanyan+2019}), as well as emission, absorption, and scattering of neutrinos in dense accretion flows around and outside the neutrinospheres~\cite{Janka+2024}. Hydrodynamic kicks are mediated by momentum transfer through asymmetric mass accretion flows and outflows, asymmetric pressure of the surrounding gas, and long-range gravitational forces between compact remnant and asymmetric ejecta~\cite{Scheck+2006}.
\begin{marginnote}[]
\entry{LESA}{Lepton Emission Self-sustained Asymmetry}
\end{marginnote}

The kicks caused by these mechanisms typically take many seconds to asymptote to their final values. Hydrodynamic kicks sometimes grow for even more than 10\,s (Figure~\ref{fig:3Dmodels}, lower right panel, and \citenum{Nakamura+2019,Janka+2022,Burrows+2024b,Janka+2024}), which confirms the importance of long-range gravitational interaction between ejecta and compact object. As expected from Eqs.~(\ref{eq:vns3}) and~(\ref{eq:vnuestimate}), SN simulations yield tight correlations of kick magnitude and the ejecta-energy and neutrino-energy dipoles $\alpha_\mathrm{ej}E_\mathrm{exp}$ and $\bar{\alpha}_{\nu}^\mathrm{tot}E_\nu^\mathrm{tot}$~\cite{Burrows+2024b,Burrows+2024}. Neutrino (LESA) induced NS kicks dominate the total kicks in ECSNe and ECSN-like events because of short post-bounce accretion phases and correspondingly early, fast, nearly spherical, and low-energy explosions. They can produce NS velocities up to $\sim$50\,km\,s$^{-1}$~\cite{Janka+2024}. NSs in explosions of more massive progenitors with longer-lasting PNS accretion can   attain neutrino kicks up to (100--150)\,km\,s$^{-1}$, but these kicks are usually subordinate to the hydrodynamic kicks, which can reach values of 1000\,km\,s$^{-1}$ and more (Figure~\ref{fig:3Dmodels}, lower right panel, and \citenum{Nakamura+2019,Mueller+2019,Janka+2022,Janka+2024,Burrows+2024b}). The total kick magnitudes thus obtained by these mechanisms are well compatible with measured and inferred kicks of radio pulsars and NSs in binary systems and associated with gaseous SN remnants. The ejecta distribution observed in many gas remnants exhibits concentrations of mass and chemical elements pointing opposite to the NS kick direction \citep{Grefenstette+2014,Grefenstette+2017,Holland-Ashford+2017,Katsuda+2018}, as expected from 3D CCSN simulations~\cite{Wongwathanarat+2013,Wongwathanarat+2017}.

Also BHs can receive kicks at their formation. Anisotropic neutrino loss mainly before the PNS collapses to the BH can cause kicks of only a few km\,s$^{-1}$, if no explosive mass ejection occurs and the entire progenitor collapses into the BH (failed SN) \cite{Janka+2024,Burrows+2024b,Burrows+2024a}, which seems compatible with the recent analysis of BH binary VFTS~243~\cite{Vigna-Gomez+2024}. However, much larger kicks can occur when the BH is formed accompanied by a successful CCSN with highly asymmetric mass ejection or later fallback of initially expelled matter that does not become gravitationally unbound, in which case hydrodynamic BH kicks may even exceed 1000\,km\,s$^{-1}$ \cite{Burrows+2024b,Burrows+2024a}, a possibility that had been predicted in Reference~\cite{Janka2013}.

New-born NSs and BHs, on the one hand, inherit the angular momentum of the stellar core or star that collapses into the compact remnant. On the other hand, however, even in nonrotating or slowly rotating stars the SASI spiral mode and asymmetric accretion can transfer significant amounts of angular momentum to the compact object, whereas the opposite angular momentum is carried away by the ejecta \cite{Blondin+2007,Rantsiou+2011,Kazeroni+2016,Kazeroni+2017}. This can produce NS spin periods ranging between $\sim$10\,ms and seconds during the onset of the explosion and shortly afterwards \cite{Wongwathanarat+2013,Mueller+2019,Stockinger+2020,Burrows+2024b,Janka+2024}, but later anisotropic accretion of fallback material with high angular momentum may change or outdo this early spin-up of the NS or BH \cite{Janka+2022,Mueller2023}. The possible spin-kick alignment inferred for some observed pulsars (e.g., \citenum{Johnston+2007,Noutsos+2012,Noutsos+2013,Yao+2021}, and references therein), if it is a generic trend of a larger population, is yet unexplained by CCSN models and controversially debated.

\begin{figure}[h]
\center
\includegraphics[width=16cm]{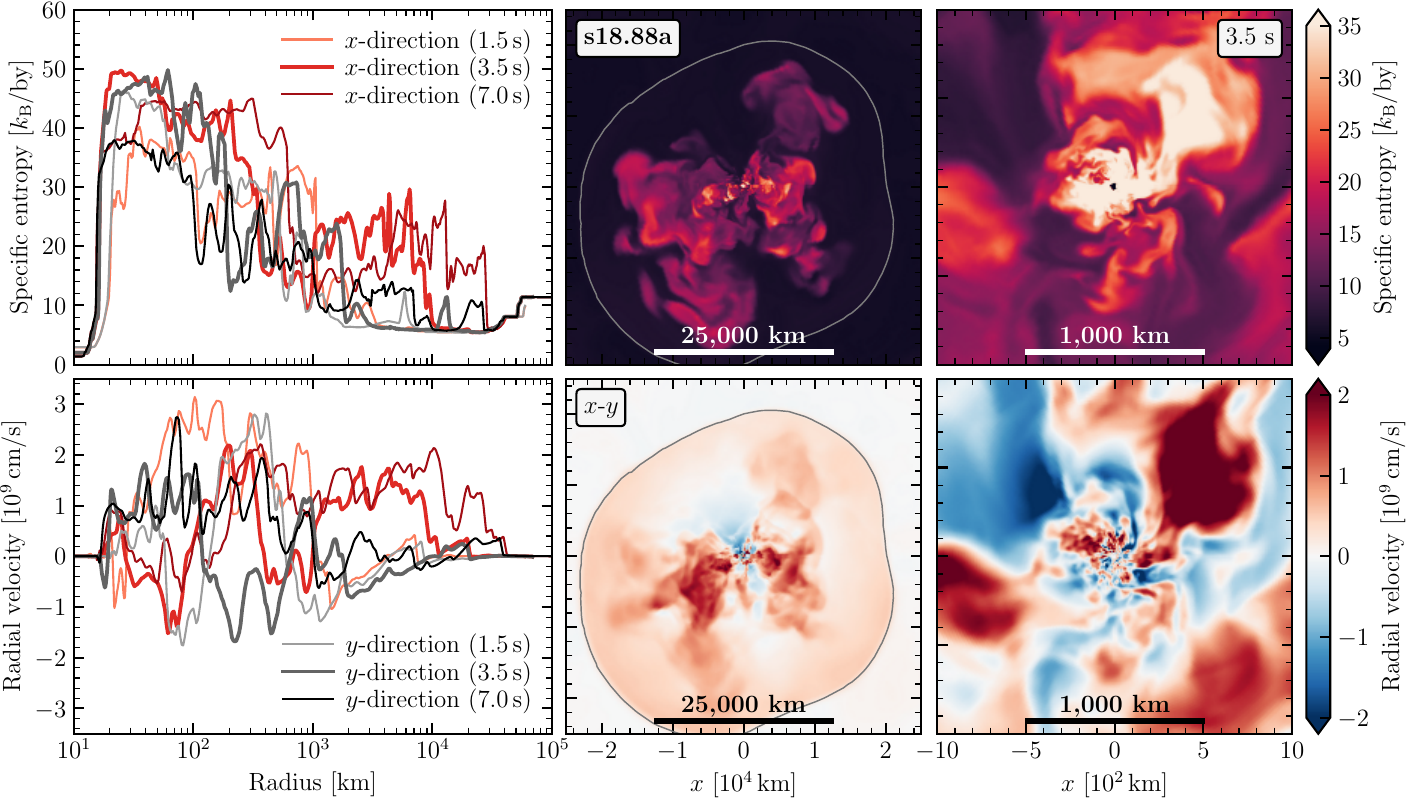}
\caption{Ejecta conditions in the long-term 3D CCSN simulation of an 18.88\,$M_\odot$ progenitor (model s18.88a of Figures~\ref{fig:3Dmodels} and \ref{fig:3Dmodels-outflows}, which is similar to model s18.88b analyzed in References \citenum{Bollig+2021,Sieverding+2023}). {\em Left panels:} Entropy per nucleon ({\em top}) and radial velocity ({\em bottom}) along randomly chosen directions (being the $x$- and $y$-directions of the cross sectional cuts shown in the middle and right panels) at 1.5\,s, 3.5\,s, and 7.0\,s after bounce. {\em Middle and right panels:} Cross-sectional cuts of the entropy per nucleon ({\em top}) and radial velocity ({\em bottom}) in the $x$-$y$-plane at 3.5\,s after bounce. The {\em middle panels} display the entire ejecta behind the deformed SN shock (thin gray line), the {\em right panels} provide blowups of the vicinity of the PNS, which is located at the center of the coordinate system. Inflows (velocities colored in blue) and outflows (velocities colored in red) lead to a highly asymmetric and unsteady ejecta distribution with secondary shocks due to multiple collisions of inflows and outflows. The neutrino-heated matter near the PNS is extremely turbulent and dynamical and provides drastically different conditions for heavy-element nucleosynthesis than the smooth and radially continuous and monotonic outflow trajectories of NDWs in 1D SN models and 3D explosions of low-mass SNe (i.e., low-mass iron-core and electron-capture SN-like events). Figure courtesy of Daniel Kresse.}
\label{fig:3Dmodels-cross-sections}
\end{figure}

\subsubsection{Neutrino-Heated Outflows and Nucleosynthesis}
\label{sec:nucleosynthesis}

NDWs are known from 1D CCSN models as neutrino-driven outflows of hot PNSs. Quasi-spherical neutrino winds occur in 3D explosion simulations only for ECSNe and ECSN-like events of low-mass stars and during short-term (intermediate or late) phases in the explosions of more massive iron-core progenitors~\cite{Stockinger+2020,Janka+2023}. The properties of these NDWs (mass-outflow rate, electron fraction, entropy, expansion time scale) depend on the radius and mass of the PNS as well as the luminosities and mean energies of the PNS's $\nu_e$ and $\bar\nu_e$ radiation~\cite{Qian+1996}. Instead of the winds, 3D explosion models show that low-entropy matter of the progenitor is pulled inward in massive accretion downdrafts that reach down toward the compact remnant for many seconds after bounce. These downflows are separated by high-entropy plumes of neutrino-heated matter that expands outward (Figure~\ref{fig:3Dmodels-cross-sections}; \citenum{Mueller+2017,Bollig+2021,Wang+2023}). 

The interaction of accretion downflows and outflows creates a highly turbulent medium in the close surroundings of the new-born NSs (Figure~\ref{fig:3Dmodels-cross-sections}), and the high mass accretion rates in the downflows lead to densities that are orders of magnitude higher than those of NDWs, whose development is therefore suppressed. For a PNS with a gravitational mass of about 1.4\,$M_\odot$, a radius of 20\,km, and a total neutrino luminosity of $6\times 10^{52}$\,erg\,s$^{-1}$, typical NDW mass loss rates are several $10^{-3}\,M_\odot$\,s$^{-1}$, and with 10\,km radius and $6\times 10^{51}$\,erg\,s$^{-1}$ they are only several $10^{-5}\,M_\odot$\,s$^{-1}$~\cite{Qian+1996,Huedepohl+2010,Huedepohl+2010a}, whereas in the 3D model s12.28 of Figures~\ref{fig:3Dmodels} and~\ref{fig:3Dmodels-outflows} with a similar PNS mass, comparing at times with the same neutrino luminosities, the inflow and outflow rates are $\sim$$2\times 10^{-2}\,M_\odot$\,s$^{-1}$ and several $10^{-3}\,M_\odot$\,s$^{-1}$, respectively. 

After the first second, the inflow and outflow rates are essentially equal in all of our 3D simulations ($\dot M_\mathrm{out}/\dot M_\mathrm{in}\sim 1$ in Figure~\ref{fig:3Dmodels-outflows}) except in the rapidly rotating model m15a. This implies that the magnitude and time evolution of the mass-outflow rates are determined by the mass infall rates from the outer layers of the progenitor rather than being defined by the properties of the PNS and its neutrino emission as in the case of NDWs. As stated in Section~\ref{sec:expergs}, the high mass inflow and outflow rates are crucial for boosting the SN explosion energies over several seconds mostly by neutrino heating with a highly time-variable effective energy input between $\sim$2\,MeV and $>$13\,MeV per nucleon (Figure~\ref{fig:3Dmodels-outflows}, bottom right; \citenum{Mueller+2017,Bollig+2021,Witt+2021}).

The nucleosynthesis conditions in the neutrino-heated outflows of 3D SN models are fundamentally different from those in (quasi-)spherical NDWs. Since the gas in the PNS surroundings is turbulent with downflows and outflows colliding with each other, secondary shocks occur and lead to a non-monotonic density, temperature, and entropy evolution of ejected mass elements~\cite{Sieverding+2023} instead of the smooth and continuous evolution of NDWs in radius and time. While 1D NDW ejecta have a single value of entropy per nucleon $s$ and electron fraction $Y_e$ at each time, describing a line in the $s$-$Y_e$-plane, outflows in 2D and 3D explosion models possess distributions of $s$ and $Y_e$ at each time and the conditions fill a complex-shaped area in this two-parameter space \cite{Stockinger+2020,Sieverding+2023,Janka+2023,Wang+2023,Wang+2024a}. The ejecta mass distributions are correspondingly wide with entropies ranging from $\sim$5\,$k_\mathrm{B}$ per nucleon up to about (70--80)\,$k_\mathrm{B}$ per nucleon and electron fractions from $\sim$0.45 or slightly lower to more than 0.6. They show considerable variations between explosions of different progenitors with partly stochastic evolution in time and possibly even case-to-case variability and dependencies on 3D perturbations in the pre-collapse progenitors. Low-mass CCSNe expel more neutron-rich matter with $Y_e$ reaching down to $\sim$0.35, connected to an early and short post-bounce phase of convective overturn \cite{Wanajo+2011,Wanajo+2018,Stockinger+2020,Wang+2024}, whereas SNe of higher-mass stars tend to produce more proton-rich ejecta \cite{Wanajo+2018,Bollig+2021,Wang+2023,Wang+2024a} and build up a high peak around $Y_e\sim$0.5 by the massive, neutrino-heated outflows that continue over many seconds. While the nucleosynthesis conditions in MD models are therefore grossly different from those obtained in 1D neutrino-driven explosions, the results of 2D and 3D CCSN simulations broadly agree in their overall behavior and basic features.

Full mass distributions and production factors of nucleosynthesis yields including special radioactive isotopes relevant for gamma-ray astronomy (e.g., $^{26}$Al, $^{60}$Fe), weak r-process nuclei, and $p$-nuclei created in the neutrino-heated ejecta during several seconds of the explosion in 3D CCSN models can be found in \cite{Sieverding+2023,Wang+2023,Wang+2024a}; see also \cite{Wanajo+2011,Wanajo+2013,Wanajo+2013a,Wanajo+2018} for similar results from a set of shorter 2D simulations with a GR hydro code and a different neutrino transport. However, simulations longer than $\sim$10\,s are needed to reach the asymptotic abundances and to avoid the need of extrapolations of the thermal histories of ejected mass elements and of the late-time mass outflow rate~\cite{Wang+2024b}. Interestingly, some radioactive isotopes (e.g., the neutron-rich nuclei $^{48}$Ca, $^{57}$Ni, $^{60}$Fe) are considerably more abundantly produced in some of the MD neutrino-driven explosions compared to 1D SN models. In particular, long-term 3D simulations solve the underproduction problem of $^{44}$Ti in 1D CCSN models~\cite{The+2006}, because neutrino-heated matter is ejected with high rates in the high-entropy outflows for many seconds (see Figures~\ref{fig:3Dmodels-outflows} and \ref{fig:3Dmodels-cross-sections}). These outflows include a major fraction of material with $Y_e\gtrsim 0.49$, which has been heated to NSE temperatures, undergoes an $\alpha$-rich freeze-out process, and experiences multiple reheating episodes with temperature increases due to collisions with secondary shocks in the turbulent PNS environment. This enables efficient $^{44}$Ti production that goes on for many seconds ($\sim$10\,s), much longer than the explosive production of $^{56}$Ni, whose yield essentially saturates within seconds. The long-lasting creation of $^{44}$Ti makes modern 3D CCSN nucleosynthesis of initially nonrotating progenitors consistent with the $^{44}$Ti and $^{56}$Ni (or iron) masses and ratios observed in SN~1987A and Cas~A \cite{Sieverding+2023,Wang+2024b}, as anticipated by 3D neutrino-driven explosion simulations in \cite{Wongwathanarat+2017}, which still employed a more approximate treatment of the neutrino physics. The $^{44}$Ti/Fe ratios for explosions of different progenitors vary moderately between several $10^{-4}$ and more than $10^{-3}$, and in the ejecta of individual explosions they span a wide range from less than $10^{-5}$ to more than 0.1, as suggested by observations of Cas~A~\cite{Grefenstette+2014,Grefenstette+2017}.

A closer inspection of the results from the 3D CCSN simulations with the \textsc{Prometheus-Vertex} and \textsc{Fornax} codes reveals interesting differences. \textsc{Fornax} does not only yield more and earlier explosions (see Section~\ref{sec:explodability}) and roughly twice as energetic ones than \textsc{Prometheus-Vertex} in comparable cases (see discussions in \citenum{Stockinger+2020,Bollig+2021,Wang+2024}), but \textsc{Fornax} simulations also yield significantly more $^{56}$Ni for similar explosion energies. For example, the 15.01\,$M_\odot$ and 16\,$M_\odot$ SN models in \cite{Wang+2024b} explode with energies of $\sim$0.3\,B and $\sim$0.4\,B and produce $\sim$0.055\,$M_\odot$ and $\sim$0.065\,$M_\odot$ of $^{56}$Ni, respectively, whereas 0.064\,$M_\odot$ of nickel are obtained in a $\sim$1\,B explosion of an 18.88\,$M_\odot$ progenitor with \textsc{Prometheus-Vertex}~\cite{Sieverding+2023}. Similarly energetic \textsc{Fornax} models (1.1--1.2\,B explosions of 18.5 and 25\,$M_\odot$ progenitors) yield $\sim$0.14\,$M_\odot$ and $\sim$0.19\,$M_\odot$ of $^{56}$Ni. Unless much of this radioactive material falls back and is accreted by the NS at times later than simulated, the \textsc{Fornax} 3D CCSN simulations of red supergiant (RSG) progenitor models seem to overproduce $^{56}$Ni compared to the average value observed for Type~IIP SNe of RSGs or blue supergiants (BSGs). Reasonable concordance between \textsc{Fornax} results and observed $^{56}$Ni masses is stated in~\cite{Burrows+2024}, but the comparison there is made with Type~IIb and~Ib/c SNe of stripped-envelope progenitors, which eject on average more than double the amount of $^{56}$Ni and iron than SNe~IIP ($0.090\pm 0.005$ and $0.097\pm 0.007$\,$M_\odot$ of $^{56}$Ni and iron, respectively, compared to $0.037\pm 0.005$ and $0.040\pm 0.005$\,$M_\odot$; \citenum{Rodriguez+2021,Rodriguez+2023}). For example, Type~IIP SN~1987A of BSG Sanduleak $-$69$^\circ$202 was diagnosed to have an explosion energy of (1--1.5)\,B and an ejected $^{56}$Ni mass of $\sim$0.07\,$M_\odot$ (e.g., \citenum{Utrobin+2021}), which is well compatible with the mentioned \textsc{Prometheus-Vertex} 18.88\,$M_\odot$ explosion.

\begin{figure}[!]
\center
\vspace{-10pt}
\includegraphics[width=15cm]{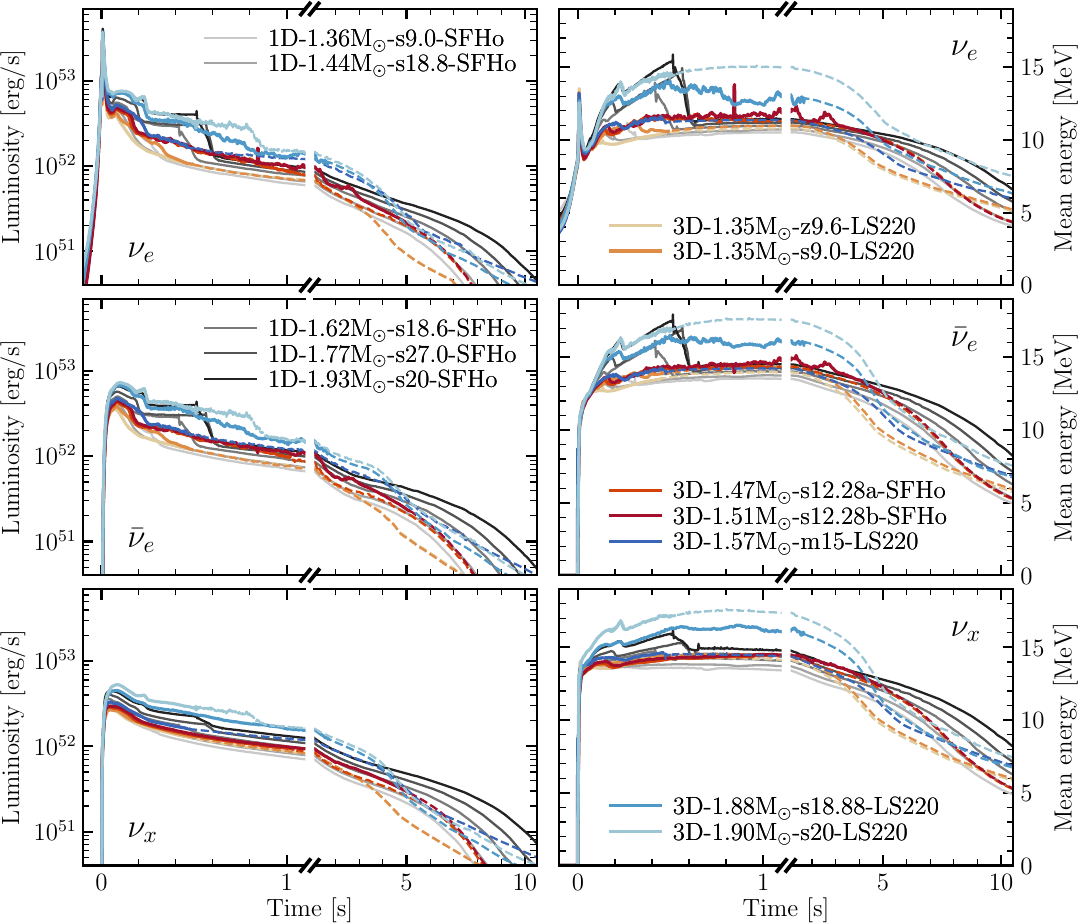}
\vspace{-5pt}
\caption{Lab-frame luminosities ({\em left}) and mean energies ({\em right}) of radiated $\nu_e$, $\bar\nu_e$, and a single species of heavy-lepton neutrinos, $\nu_x$, vs.\ post-bounce time with the well-known phases of prompt $\nu_e$ burst (duration $\sim$10\,ms), post-bounce accretion ($\sim$0.1--1\,s), and PNS cooling (seconds). A set of 3D SN simulations with different PNS masses and different nuclear EoSs is compared with 1D cooling simulations for PNSs with similar masses and the SFHo EoS (Kresse et al., in preparation). The 1D calculations (thin gray and black lines) were done with the \textsc{Prometheus-Vertex} neutrino-hydrodynamics code and MLT PNS convection. The 3D simulations (bold colored solid lines) employed \textsc{Vertex} neutrino transport for 0.5--5.1\,s and used the \textsc{Nemesis} scheme afterwards (colored dashed lines) with \textsc{Vertex} applied for the longest evolution times in model s12.28b (5.1\,s) and models s18.88a,b (1.7\,s). The 3D \textsc{Vertex} results are angle-averaged. The model names listed in the different panels (for 1D {\em left}, 3D {\em right}) are composed of dimension, baryonic PNS mass, explosion model with progenitor ZAMS mass, and employed nuclear EoS. 3D models exhibit features connected to long-lasting PNS accretion, whereas 1D models show the characteristic step-like decline of the neutrino luminosities and mean energies after the artificial initiation of the SN explosion. Overall, however, the 3D neutrino signals show the same basic features as the 1D results and closely follow the time evolution of the 1D cases for similar PNS masses and EoSs, because the 1D calculations include MLT PNS convection. During the post-bounce accretion phase the $\nu_e$ and $\bar\nu_e$ luminosities and mean energies are determined by accretion emission, which sensitively depends on the PNS mass but less strongly on the nuclear EoS. In contrast, both PNS mass and nuclear EoS have an impact on the $\nu_x$ luminosity and mean energy in the accretion phase. During the long-term evolution after 1\,s, the nuclear EoS is most crucial for the decline behavior of the cooling luminosities, and the PNS mass determines the magnitude and signal duration. Note that all 1D calculations were done with the SFHo EoS, whereas most 3D models (except two) used LS220 (for a detailed discussion of the EoS effects in 1D models with MLT convection, see Reference~\citenum{Lucente+2024}). Figure courtesy of Daniel Kresse.}
\label{fig:neutrino-signals}
\end{figure}

\subsubsection{Neutrino Signals}

The elementary phases and general properties of the neutrino luminosities and spectra radiated by CCSNe and nascent NSs have been described in numerous recent reviews, e.g., \cite{Mirizzi+2016,Janka2017a,Mueller2019}. Figure~\ref{fig:neutrino-signals} displays results obtained with the \textsc{Prometheus-Vertex} code for sets of 1D and 3D simulations of SN explosions and PNS cooling with different nuclear EoSs.

Neutrino signals from a large set of long-term 2D CCSN simulations, based on one of the currently most elaborate treatments of neutrino transport, have meanwhile been published \cite{Nagakura+2021a,Vartanyan+2023a} as well as a smaller sample of 3D results with a few cases also extending to several seconds post-bounce~\cite{Bollig+2021,Nagakura+2021b,Vartanyan+2023a,Janka+2024}.
However, there are significant differences in the neutrino emission properties predicted by modern SN models using different neutrino transport codes and neutrino interactions. This is not too astonishing in consideration of the quantitative differences of MD CCSN results mentioned in Section~\ref{sec:nucleosynthesis} and the qualitatively different outcomes discussed in Section~\ref{sec:explodability}. For example, \textsc{Fornax} yields a 3D explosion for a 40\,$M_\odot$ progenitor model~\cite{Burrows+2023}, whereas \textsc{Prometheus-Vertex} does not for the same progenitor~\cite{Janka+2024}. Some indications of significant differences in the neutrino signals can indeed be seen in the 1D comparison of~\cite{OConnor+2018c}, although based on a test setup with reduced and controlled microphysics inputs. 

Neutrino signals from 2D simulations display time variability and fluctuations with substantially overestimated amplitudes and for too extended periods of time, because the artificial constraint of axisymmetry favors the existence of long-lasting and very massive equatorial accretion flows, whose impact and settling on the PNS surface leads to enhanced accretion luminosities. These equatorial flows are actually toroidal sheets, which separate plumes of neutrino-heated ejecta expanding toward both polar directions. Because of the interaction of downflows and outflows, the equatorial accretion sheets are highly time variable, thus fueling strong and irregular mini-bursts of neutrino emission~\cite{Mueller+2014}. The long-time, angle-averaged neutrino signals of 3D simulations are usually smoother and exhibit fewer and less extreme eruptions after the post-explosion accretion by the PNS has abated at typically $\sim$1--3\,s, depending on the model (see the 3D results in \citenum{Vartanyan+2023a,Janka+2024}). Such accretion features can be witnessed in models s12.28b and s18.88 in Figure~\ref{fig:neutrino-signals}, which were started from 3D pre-collapse progenitor data and evolved with \textsc{Vertex} neutrino transport until 5.1\,s and 1.7\,s post-bounce, respectively. The reason for the 2D/3D differences is twofold. On the one hand, the downflows in 3D are less massive and more numerous, because they are tube-like accretion funnels rather than toroidal sheets. This implies less extreme accretion fluctuations but also a higher degree of angular averaging. On the other hand, in 3D a larger fraction of the matter is blown out again after the gas has absorbed energy from neutrinos and before it is accreted onto the PNS~\cite{Mueller+2015}. The 2D geometry impedes the re-ejection of the infalling matter, which leads to a long-lasting, significant increase of the PNS mass in 2D, whereas the enhanced re-ejection of neutrino-heated downflow material causes a continuous and prolonged growth of the explosion energy and more energetic explosions in 3D. 

In fact, when the phase of massive PNS accretion before and after the onset of the explosion is over, the neutrino signal properties of 3D simulations tally with those of 1D cooling models for similar PNS masses, if PNS convection by an MLT treatment is taken into account (see Figures~\ref{fig:PNSconvection} and~\ref{fig:neutrino-signals}). In 1D simulations PNS accretion ends quite abruptly when the explosion is artificially triggered and the shock expansion cuts off further mass infall. Therefore the luminosities decline from the accretion plateau to the PNS cooling tail quite sharply within only a few 100\,ms. In contrast, in 3D simulations significant mass accretion by the PNS can continue for several seconds after the onset of shock expansion (though more weakly than in 2D simulations, as mentioned above) and accretion emission enhances mainly the $\nu_e$ and $\bar\nu_e$ luminosities for up to $\sim$1\,s. Afterwards the radiated luminosities and mean energies transition to the PNS cooling behavior more gradually than in 1D, but they asymptote very closely to the level of 1D models for a comparable PNS mass. The slope of the late-time decline of the PNS cooling emission is determined by the nuclear EoS and the symmetry-energy dependent evolution of PNS convection~\cite{Roberts+2012}, while the magnitude and duration of the emission is governed by the NS mass (Figure~\ref{fig:neutrino-signals}). A systematic analysis of the long-term evolution of neutrino luminosities for PNS models with different masses and different nuclear EoSs including PNS convection by an MLT treatment can be found in~\cite{Lucente+2024}.  

Although accretion by the PNS never completely stops but persists episodically, the accretion rates are usually so low (reaching at most some $10^{-3}\,M_\odot$\,s$^{-1}$ in sub-second long peaks, because inflow and outflow rates are essentially balanced; $\dot M_\mathrm{out}/\dot M_\mathrm{in} \sim 1$ in Figure~\ref{fig:3Dmodels-outflows}) that the accretion emission adds relatively little to the neutrino radiation produced by the cooling PNS. Only at very late times (later than $\sim$7\,s post-bounce, model dependent), when the PNS emission has decayed and therefore the diminishing neutrino heating gives way to fallback accretion onto the PNS ($\dot M_\mathrm{out}/\dot M_\mathrm{in} < 1$ develops), the fallback emission can become the dominant source of neutrinos. However, this fallback accretion does not take place from a spherical flow, creating a steady-state, shocked accretion mantle of the PNS as assumed in~\cite{Akaho+2024}. Instead, the PNS environment is turbulent and the highly dynamical conditions visible in Figure~\ref{fig:3Dmodels-cross-sections} with colliding and mixing inflows and outflows exist during all of the simulated evolution until well beyond 10\,s. The longest 3D simulations performed so far, reaching nearly 20\,s, show a continuous (but nonmonotonic) trend of decrease in $\dot M_\mathrm{out}/\dot M_\mathrm{in}$. Because of turbulent mixing of low-entropy and high-entropy matter, the medium in the immediate surroundings of the PNS ($r\lesssim 100$\,km) is nearly isentropic with entropies of several 10\,$k_\mathrm{B}$ per nucleon. Therefore the angle-averaged temperature and density profiles can be well approximated by power laws, $T(r)\approx T_0(R_\mathrm{NS}/r)$ and $\rho(r)\approx \rho_0(R_\mathrm{NS}/r)^3$. A lot of the infalling matter, however, does not reach the close vicinity of the PNS and the narrow shell (radial thickness $\sim$5\,km) of efficient neutrino cooling. Due to its high angular momentum and large total energy, much of the fallback gas passes the PNS and returns to greater radii, filling a bigger volume around the PNS. Since the angular momentum is likely to delay accretion, this matter will be able to settle onto the PNS only over much longer time scales, or it will ultimately escape from the gravitational attraction of the new-born NS with low expansion velocities or accelerated by magnetic activity of the new-born NS. All current long-term 3D CCSN models keep the NS fixed at the origin of the computational grid and thus ignore the movement of the kicked NS (see Section~\ref{sec:kicks}). However, the gradual displacement of the kicked NS from the center of the explosion is likely to play an important role when considering the interaction of the newly formed compact remnant with its fallback environment over even longer evolution times; this poses a generic 3D problem to be tackled by future work~\cite{Janka+2022}.

Several recent papers have addressed the important question whether the neutrino signals predicted by modern SN models are compatible with the detected SN~1987A neutrinos~\cite{Olsen+2021,Olsen+2022,Fiorillo+2023,Li+2024}. Considering a set of long-term 1D cooling simulations for PNSs with systematically varied masses and EoSs, obtained from artificially triggered SN explosions and computed with the \textsc{Prometheus-Vertex} code including MLT convection, Reference~\cite{Fiorillo+2023} reported overall agreement of the total energy and mean neutrino energy with the joint 95\% confidence regions for all experiments [Kamiokande~II (Kam-II), Irvine-Michigan-Brookhaven (IMB), Baksan Underground Scintillator Telescope (BUST), and Liquid Scintillator Detector (LSD)]; for a determination of best-fit parameters, see also \cite{Olsen+2021,Olsen+2022,Bozza+2025}. Moreover, the modeled signal duration agrees well with the IMB burst. However, the predicted signals are too short to explain the last three events between 9.2\,s and 12.4\,s in Kam-II and the last two events in BUST (at 7.7\,s and 9.1\,s), mainly because PNS convection shortens the neutrino emission from Kelvin-Helmholtz cooling to $\lesssim$10\,s, at which time the luminosities for each neutrino species have fallen well below $10^{51}$\,erg\,s$^{-1}$ even for the most massive considered NS and for all employed EoSs (Figure~\ref{fig:neutrino-signals}). In \cite{Li+2021} a tension of number counts and especially mean energies was claimed to exist between models and data during the first second, but this study employed an inhomogeneous set of 1D and MD models, which partly did not explode and applied neutrino transport and rates with different detailedness. Moreover, angle-averaged neutrino results from MD models were used and therefore variations with the viewing direction (up to several 10\% for the luminosities in 3D;~\citenum{Vartanyan+2019}) were not taken into account. The tension was not confirmed by the analysis in~\cite{Fiorillo+2023}, where the statistical significance of a mismatch between model spectra and detected event energies was attributed to a local fluctuation of the data, because it disappears again when longer data intervals are considered.

The short PNS cooling signals require an alternative explanation of the latest events in Kam-II and BUST. Additional neutrino emission from fallback accretion is an interesting possibility (\citenum{Fryer+2009}; see also \citenum{Akaho+2024}). Indeed, our 3D simulations show considerable time variability of the mass infall rate with stochastic excursions, and in model s18.88b a large and seconds-long increase of $\dot M_\mathrm{in}$ peaking near $10^{-2}\,M_\odot$\,s$^{-1}$ can be witnessed between $\sim$9\,s and $\sim$12\,s, which is connected to a massive downflow in one hemisphere that channels several $10^{-3}\,M_\odot$ toward the PNS within this time interval. Therefore fallback accretion luminosities of 
\begin{equation}
   L_{\nu,\mathrm{acc}} \sim \frac{GM_\mathrm{NS}\dot M_\mathrm{acc}}{R_\mathrm{NS}} \sim 4.5\times 10^{51}\left(\frac{M_\mathrm{NS}}{2\,M_\odot}\right) \! \left(\frac{\dot M_\mathrm{acc}}{0.01\,M_\odot/\mathrm{s}}\right) \! \left(\frac{R_\mathrm{NS}}{12\,\mathrm{km}}\right)^{\!\!-1}\ \mathrm{\frac{erg}{s}}
   \label{eq:lacc}
\end{equation}
seem possible in extreme cases and are mostly shared by $\nu_e$ and $\bar\nu_e$ produced via charged-current reactions in the hot, turbulent medium around the PNS. Such accretion luminosities dominate the PNS emission at around 10\,s after bounce by far (Figure~\ref{fig:neutrino-signals}) and are much larger than the fallback luminosities expected in~\cite{Li+2021}. Neutrino emission from a fallback episode like the one in model s18.88b is already marginally compatible with the 95\% confidence region for the events detected between $\sim$9\,s and $\sim$12\,s in Kam-II (Figure~11 in Reference~\citenum{Fiorillo+2023}). Note that this late neutrino emission is not visible in Figure~\ref{fig:neutrino-signals}, because the version of the \textsc{Nemesis} neutrino approximation applied in the 3D simulations up to 20\,s did not yet include the relevant physics. 

Alternatively to the fallback scenario, the explanation of the latest SN~1987A neutrino events might be connected to the nuclear EoS. Either the properties of the EoS in the super-nuclear core prevent convection from becoming as strong as in the current models and thus it advances inward only much more slowly, thereby delaying the convective transport of energy and electron lepton number out of the PNS core. Or a late phase transition, for example from hadrons to quark matter, releases internal or gravitational energy (due to a second collapse) and reheats the compact remnant (e.g., \citenum{Fischer+2018,Jakobus+2022}), which should extend the neutrino-cooling signal for several seconds.

\begin{textbox}[h]\section{NEUTRINO-FLAVOR CONVERSION}
One of the major uncertainties in current CCSN modeling is the occurrence and treatment of neutrino flavor conversion in the SN core, in particular between the neutrino-decoupling region around the neutrinospheres and the gain layer where neutrinos deposit energy driving the SN shock. Only in recent years the possibility and importance of so-called fast pair-wise neutrino oscillations in this region has been recognized~\cite{Izaguirre+2017,Bhattacharyya+2021}. Fast flavor conversion (FFC) is a self-induced phenomenon connected to $\nu$-$\nu$ refraction, spawning this collective behavior, which depends on the density and angular distribution of the neutrinos. It takes place on small length and time scales (cm and ns) and therefore it cannot be resolved by global MD SN simulations with a spatial resolution of typically $\gtrsim 100$\,m and time steps of $\sim$10$^{-7}$\,s. Moreover, since FFC is a multi-angle effect, direct calculations require the full phase space distribution of the neutrinos, which is not available in current full-scale and long-term SN calculations. Effective treatments of FFC effects on large scales are needed, based on a detailed understanding of the physics on the unresolved scales. So far only first steps have been taken to investigate the impact of FFC in dynamical SN models. These studies reveal a possible influence on the detectable neutrino signal, the neutrino-driven explosion mechanism, SN nucleosynthesis, new features in the emitted GW signal, and possibly neutrino-induced SN kicks \cite{Ehring+2023a,Ehring+2023b,Nagakura2023,Ehring+2024,Nagakura+2024,Wang+2025}. The enormous interest in this fascinating phenomenon has led to rapid developments propelled by a large community of active researchers. The corresponding accomplishments and status cannot be adequately reported here and the reader is referred to dedicated papers and the references there~\cite{Abbar+2019,Tamborra+2021,Nagakura+2021,George+2024,Xiong+2024,Johns+2025}.
\end{textbox}

\begin{figure}[h]
\center
\begin{minipage}[b]{6.0cm}
\includegraphics[width=0.98\textwidth]{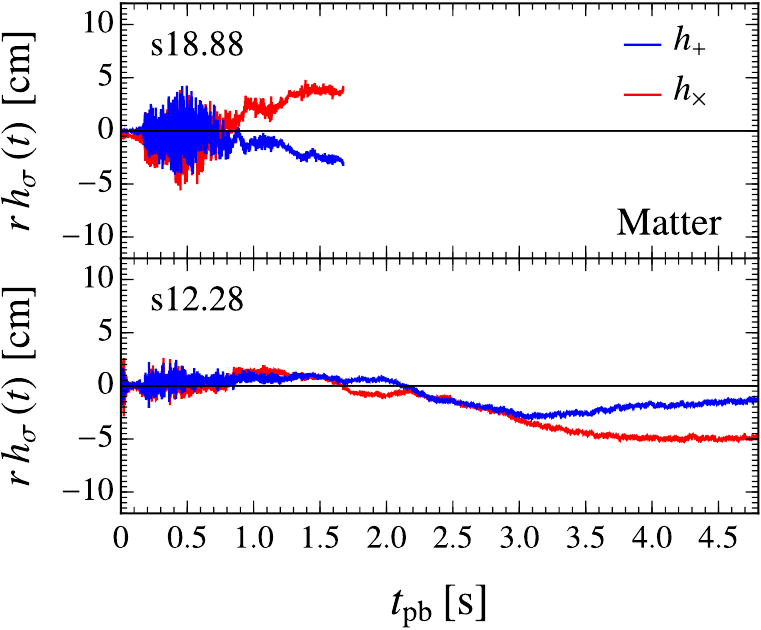}
\end{minipage}\hspace{5pt}
\begin{minipage}[b]{6.0cm}
\includegraphics[width=1.0\textwidth]{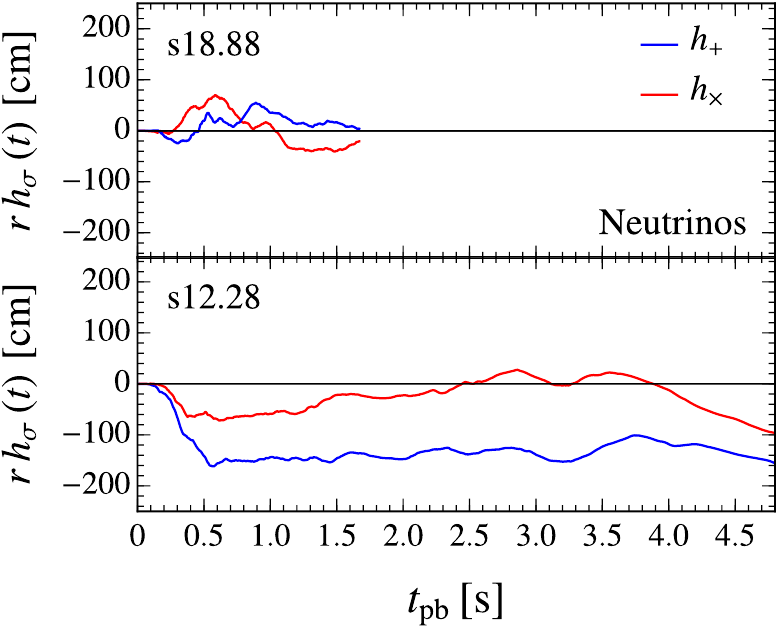}
\end{minipage}\\\vspace{-10pt}
\begin{minipage}[b]{6.0cm}
\includegraphics[width=1.0\textwidth]{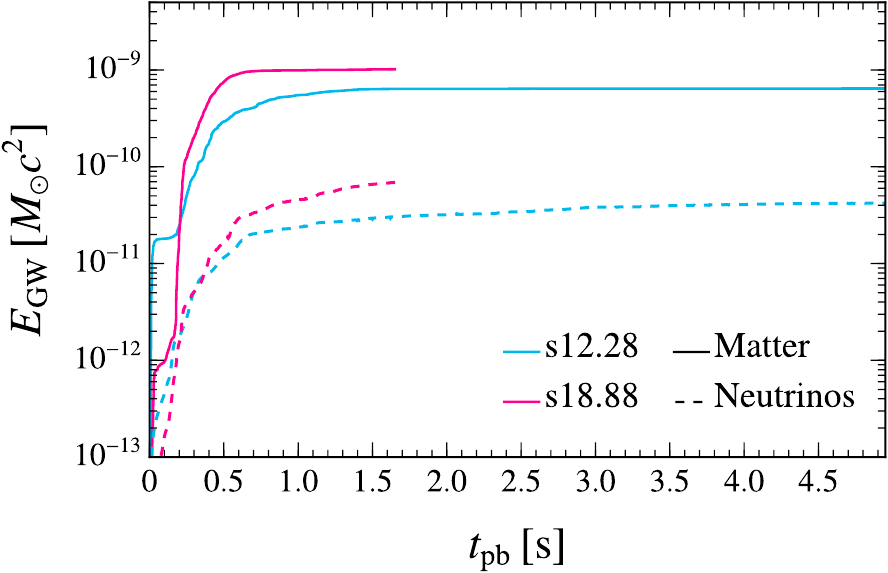}
\end{minipage}\hspace{5pt}
\begin{minipage}[b]{6.0cm}
\includegraphics[width=1.0\textwidth]{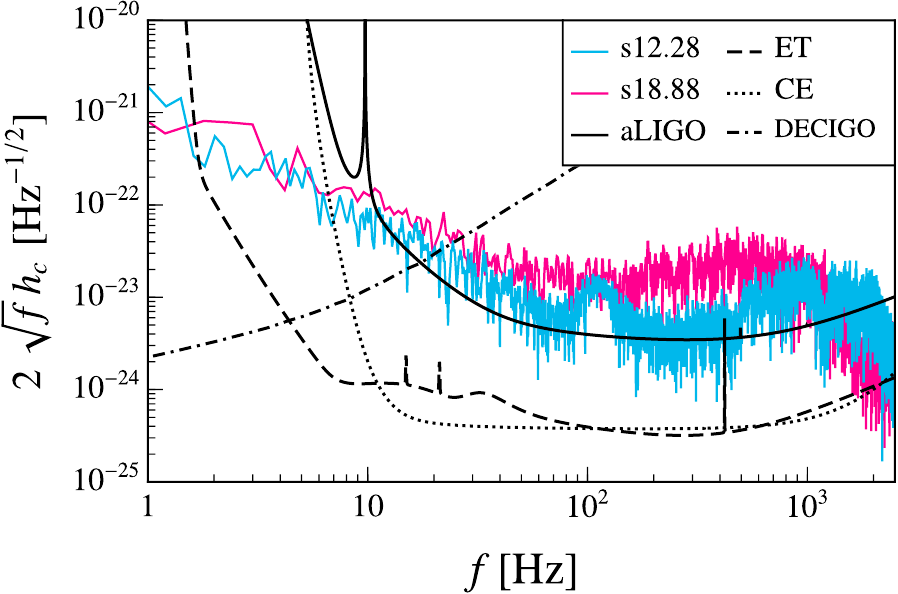}
\end{minipage}
\caption{GW signals produced by asymmetric mass motions and anisotropic neutrino emission in 3D explosions of 12.28\,$M_\odot$ and 18.88\,$M_\odot$ progenitors \citep{Sukhbold+2018} (explosion models s12.28b and s18.88b of Figure~\ref{fig:3Dmodels}), post-processed by A.~Lella et al.~\cite{Lella+2026}. The CCSN simulations were carried out with the \textsc{Prometheus-Vertex} code (R.~Bollig, private communication, and References~\citenum{Bollig+2021,Janka+2023}) for progenitors with convective oxygen-shell burning computed in 3D (during the final hour before CC in the 12.28\,$M_\odot$ case and the final 10\,min in the 18.88\,$M_\odot$ case; Reference~\citenum{Yadav+2020}). {\em Top:} GW amplitudes for the plus and cross polarizations of the matter signals ({\em left}) and the total neutrino signals (sum for all neutrino species; {\em right}) as functions of post-bounce time $t_\mathrm{pb}$ for an observer located in the equatorial plane of the source. {\em Bottom left:} Cumulative GW energies for the matter signals (solid lines) and neutrino signals (dashed lines) of both SN models. {\em Bottom right:} Amplitude spectral densities as functions of frequency for the combined neutrino and matter GW signals of both SN models assuming a galactic SN at a distance of 10\,kpc. The black lines depict the sensitivity curves of current (Advanced LIGO, \citenum{advLIGO-2015}) and forthcoming GW experiments (Einstein Telescope, \citenum{ET-2010a,ET-2010b}; Cosmic Explorer, \citenum{CE-2017}; DECIGO, \citenum{DECIGO-2008}) operating in the frequency range of interest. Note that in the case of model~s18.88 the GW power at frequencies above $\sim$1000\,Hz is underestimated because of sparse output sampling with time intervals of 0.5\,ms. Figure courtesy of Alessandro Lella.}
\label{fig:GWsignals}
\end{figure}

\subsubsection{Gravitational Waves}

Alike neutrinos GWs are an important direct probe of the processes taking place in the SN core. They are generated by temporal changes of the mass/energy-quadrupole moment and are thus produced by asymmetric mass motions as well as anisotropic neutrino emission. A growing number of MD simulations reveals that the GW signals from CCSNe are stochastic, displaying enormous variability in time, viewing direction, and from case to case. However, there are well identified signal components and various systematic trends (for reviews, see \citenum{Ott2009, Kotake2013, KotakeKuroda2017_Handbook, Kalogera+2021, Abdikamalov+2022, Mezzacappa+2024}).

Collapsing stellar cores produce a characteristic millisecond-long GW pulse at the moment of core bounce, if they get deformed because of sufficiently rapid rotation, and later on they can radiate strong GW signals due to spiral modes, triaxial hydrodynamic (spiral and bar-mode) and magnetohydrodynamic instabilities, mass ejection in polar jets or, in the most extreme case, even off-center density maxima and core fragmentation, all connected to rapid rotation. But also the collapse of nonrotating stars can act as a source of GWs, because asymmetries develop by hydrodynamic instabilities such as prompt post-bounce convection (for tens of ms) due to a negative entropy gradient behind the decelerating core-bounce shock, violent SASI and convective overturn in the neutrino-heated postshock layer (for hundreds of ms), and PNS oscillations (g-, p-, and f-mode activity for many seconds), stimulated mainly by accretion downdrafts hitting the NS from outside and less strongly by convection inside the PNS (e.g., \citenum{Andresen+2017,Radice+2019,Torres-Forne+2019,Torres-Forne+2021,Powell+2020,Andresen+2021,Vartanyan+2023,Choi+2024,Powell+2024}). After the onset of the explosion, GW emission continues from long-lasting PNS convection and vibrations, fallback accretion, and the asymmetric morphology of the expanding explosion ejecta and radiated neutrinos, both leading to a long-term GW strain or ``memory'' \cite{Braginsky+1987,Christodoulou1991,Favata2010}.
\begin{marginnote}[]
\entry{GW ``memory''}{Abiding systematic offset of the GW amplitude from the zero-level due to anisotropic neutrino emission and aspherical explosive mass ejection}
\end{marginnote}

Typical distance-independent GW quadrupole amplitudes obtained in 3D simulations are on the order of several cm for the matter-associated signal (Figure~\ref{fig:GWsignals}, upper left; \citenum{Kuroda+2016a,Kuroda+2017,Andresen+2017,Radice+2019,Powell+2019,Vartanyan+2019,Mezzacappa+2023,Vartanyan+2023,Choi+2024}, which is a factor 10--20 smaller than in 2D simulations~\cite{Mueller+2013,Yakunin+2015}, but for very massive and rapidly rotating progenitors peak amplitudes of several 10\,cm can also be obtained in 3D \cite{Powell+2020,Powell+2024}. (The detector strains require division by the source distance $D$ in 3D and by $3.66D$ for optimal source orientation in 2D.) Although the GW amplitudes are overestimated in 2D, the basic signal features and associated frequencies are well comparable with 3D results. The matter signal covers a wide range of frequencies from tens of Hz up to several 1000\,Hz, in particular before and shortly after the onset of the explosion. Hydrodynamic instabilities in the postshock layer (convection, SASI) cause GW emission mainly around 100--200\,Hz (twice the SASI frequency), whereas oscillations and pulsations of the PNS, which are mostly instigated in its near-surface layers by the impact of accretion downdrafts, produce high-frequency, broad-band GW emission, whose dominant frequency characteristically increases from a few 100\,Hz shortly after bounce to more than 1000\,Hz after about a second, signaling the contraction of the PNS due to neutrino losses. In exploding models this high-frequency component transitions into a narrow-band signal (assigned to the fundamental PNS f-mode quadrupolar oscillation) with further increasing frequency reaching up to more than 2000\,Hz in low-mass NSs and over 3000\,Hz in high-mass NSs after several seconds. Moreover, exploding models display a strong low-frequency ($\lesssim$50\,Hz) contribution from the matter memory of the asymmetrically expanding ejecta (visible as the gradually growing offset of the matter amplitudes from the zero-level in Figure~\ref{fig:GWsignals}, upper left panel). 

The exploding models in Figure~\ref{fig:GWsignals} ($E_\mathrm{exp} \sim 0.25$\,B and 1.0\,B, respectively) radiate GW energies of up to $10^{-9}$\,$M_\odot c^2$ (lower left panel), but values up to $10^{-7}$\,$M_\odot c^2$ were reported in~\cite{Choi+2024} for very massive and exploding progenitors, and even higher values seem possible with rapid rotation~\cite{Powell+2024}. A direct scaling of the GW amplitudes and energy with the convective (or turbulent) energy in the postshock layer and thus with the explosion energy was predicted~\cite{Mueller2017,Powell+2019} and was confirmed by a growing collection of 3D CCSN simulations~\cite{Radice+2019}, which also revealed a positive correlation with the compactness of the progenitor's core~\cite{Vartanyan+2023}, which in turn correlates with the explosion energy~\cite{Nakamura+2015,Burrows+2024}. 

The neutrino memory signal dominates the GW power mainly in the low-frequency domain ($\lesssim$10\,Hz) with amplitudes up to $\sim$200\,cm, but because of the low frequencies it contributes to the total radiated GW energy at least an order of magnitude less than the matter signal (Figure~\ref{fig:GWsignals}, upper right and lower left panels). In extreme cases of massive progenitors with highly asymmetric neutrino emission and/or very large radiated neutrino energies, neutrino-GW amplitudes up to $\sim$2000\,cm and corresponding energies of several $10^{-10}$\,$M_\odot c^2$ were witnessed with still growing values after seconds of post-bounce evolution \cite{Vartanyan+2023,Choi+2024}.   

A detailed GW measurement from a future galactic CCSN will permit one to diagnose various hydrodynamic and otherwise inaccessible effects in the cores of collapsing stars, e.g., rapid rotation, the contraction of the hot PNS, stochastic and quasi-periodic (SASI) shock motions, the onset time of the explosion, PNS pulsation modes affected by the nuclear EoS, and the asymmetric morphology of the inner ejecta. However, quantitatively reliable predictions of the GW properties require 3D models with full GR, because Newtonian simulations underestimate the GW frequencies, and the approximate inclusion of GR effects in the majority of current 3D CCSN models tends to overestimate them (by $\sim$20\%;~\citenum{Mueller+2013}).

\subsubsection{Black Hole Formation}

Along with the remarkable progress in long-term MD SN modeling, special focus has also been put on BH formation in collapsing stars and has led to new insights, in particular that BH formation can be accompanied by successful explosions in various ways. Shock expansion before the PNS collapsed to a BH had been seen already in previous 2D and 3D CC simulations (e.g., \citenum{Pan+2018,Kuroda+2018,Pan+2021,Powell+2021,Rahman+2022,Janka+2024}), but either these simulations stopped at a point where they were still inconclusive about mass ejection, or the shocks were too weak to trigger any powerful, SN-like explosion. In \cite{Chan+2018,Chan+2020} a zero-metallicity (Population~III) progenitor model was collapsed and tuned such that delayed shock revival and (ultimately weak) explosions were triggered by neutrino heating due to intense luminosities and hard spectra of $\nu_e$ and $\bar\nu_e$, before BHs formed. The subsequent growth of the BH mass by continued accretion of fallback matter was tracked until shock breakout at the stellar surface.
\begin{marginnote}[]
\entry{Metallicity}{Abundance of ``metals'', i.e., of elements heavier than helium, here in the plasma of stars at the time of their birth}
\end{marginnote}
\begin{marginnote}[]
\entry{Population~III stars}{First generation of possibly very massive stars with zero metallicity in the very early universe}
\end{marginnote}

Recently, self-consistent explosions of this kind with a range of explosion energies and BH formation via accretion over time scales of $\sim$100\,ms to $\sim$2\,s after the shock revival were witnessed for a set of eight zero-metallicity progenitors with masses between 60\,$M_\odot$ and 95\,$M_\odot$~\cite{Sykes+2024a}. The corresponding 2D simulations also followed the blast-wave evolution and fallback accretion by the BH until shock breakout. Similarly, in \cite{Burrows+2023,Burrows+2024a} such a case of BH formation accompanied by a powerful explosion was obtained in a fully self-consistent 3D collapse simulation of a solar-metallicity 40\,$M_\odot$ model with the physics used in the \textsc{Fornax} code. The same progenitor, however, collapses ``silently'' to a BH when simulated in 3D with the \textsc{Prometheus-Vertex} code, i.e., the BH is assembled in a ``direct'' collapse through uninterrupted accretion of infalling stellar matter onto the transiently stable NS, while the shock recedes (see~\citenum{Janka+2024}). 

The BH formation time in such CCSNe (whose naming is still ambiguous; in Reference~\citenum{Sykes+2024a} they are called ``fallback SNe'', whereas Reference~\citenum{EggenbergerAndersen+2024} terms them ``BH-SNe'') as well as the explosion energy, ejecta and compact-remnant masses, and nucleosynthesis including the amount of expelled $^{56}$Ni depend on the nuclear EoS (e.g., \citenum{OConnor+2011,Steiner+2013,Powell+2021}); in~\cite{EggenbergerAndersen+2024} these dependencies were explored by 2D long-term simulations using Skyrme-type EoSs with different values of the effective nucleon mass to probe the impact of varied thermal contributions to the pressure of the nucleon gas. A higher nucleonic effective mass reduces this thermal pressure and triggers a faster contraction of the PNS and its earlier collapse to a BH. The early BH formation shortens the phase of neutrino-energy transfer to the shock and weakens or prevents an explosion. (Explosion stochasticity due to axis artifacts might be the reason why this correlation is not perfect in the 2D models of Reference~\citenum{EggenbergerAndersen+2024}.) Any other effect that softens the EoS and/or reduces the maximum mass of hot (or cold) NSs, e.g., muon creation in the PNS medium, can have similar consequences and can make an explosion less likely.  

Since post-bounce PNS accretion in 2D and 3D CCSN models is a long-lasting phenomenon or does not completely stop (except for short, irregular periods of time and in explosions of ECSNe and some low-mass iron-core progenitors, where NDWs can develop) before neutrino heating abates and fallback sets in, the huge variation of the density structures of progenitors allows for a diversity of BH formation scenarios. The progenitor's density profile determines the initial mass and accretion rate of the PNS. If this accretion proceeds with a higher rate, the PNS as a neutrino source survives for a shorter time, reducing the neutrino-energy deposition that powers the neutrino-driven explosion. 

\begin{marginnote}[]
\entry{PPI, PPISN}{Pulsational pair-instability means a violent pulsation of a very massive star ($\sim$70--140\,$M_\odot$), triggered by $e^+e^-$-pair formation in the stellar core, leading to SN-like mass ejection (PPISN)}
\end{marginnote}
Based on their large set of long-term 3D CCSN simulations, Burrows et al.~\cite{Burrows+2024a} witnessed different paths to BH formation and distinguished four corresponding ``channels'', ordered by decreasing strength of the neutrino-powered shock: Channel~1, where BHs from the maximum NS mass upward are formed within seconds by continuous mass accretion in stars (19.56\,$M_\odot$ and 40\,$M_\odot$ in the model set) that still explode, potentially strongly ($\gtrsim$1.75\,B), as described above; Channel~2, where a relatively weak explosion ($\lesssim$0.5\,B for a 23\,$M_\odot$ model) leaves a NS that collapses to a BH by fallback on time scales of tens to hundreds of seconds (corresponding to what was originally understood as ``fallback SN'' and was also predicted on grounds of 1D explosion models with neutrino engines, e.g., \citenum{Sukhbold+2016,Mueller+2016a,Ertl+2020}); Channel~3, where a very massive pre-collapse star with pulsational pair-instability (PPI) history (100\,$M_\odot$ in Reference~\citenum{Burrows+2024a}, but similar results were found for Population~III and PPI progenitors with ZAMS masses between 60\,$M_\odot$ and 115\,$M_\odot$ in References \citenum{Powell+2021,Rahman+2022,Sykes+2024a,Janka+2024}) collapses to a BH within fractions of a second and the revived shock is too weak to initiate a SN explosion and to prevent the collapse of essentially the entire progenitor; and Channel~4, where the stalled bounce shock is never revived but the stellar collapse (for 12.25\,$M_\odot$ and 14\,$M_\odot$ cases in the model set of Reference~\citenum{Burrows+2024a}) leads straightaway to a BH (this is the BH birth in a failed SN imagined as a ``quiescent'' or ``direct'' implosion of a star).    

These formation paths of stellar-mass BHs and possibly additional variants (e.g., connected to a phase transition at super-nuclear densities) may exist, depending on the EoS and accretion-rate dependent survival time of the PNS and the energy transfer by neutrinos to the surrounding, infalling stellar gas. However, the association with progenitor masses in a given pool of stellar models must still be considered as highly uncertain because of the repeatedly addressed differences in the CC outcomes obtained with different SN codes (e.g., Section~\ref{sec:explodability}).

\begin{summary}[SUMMARY POINTS]
\begin{enumerate}
\item The viability of the neutrino-driven mechanism is supported by a growing number of MD neutrino-hydrodynamics simulations. However, different state-of-the-art codes do not agree in their predictions of SN explosions and explosion properties or BH formation of individual progenitors. The explosion systematics of massive stars depending on the stellar mass are therefore still controversial.
\item Explosion energies of CCSNe can take several seconds to reach their terminal values, persistently fueled by neutrino energy transfer to matter that is accreted from the progenitor in an inflow-outflow cycle continuing for many seconds.
\item Long-term 3D simulations show that NDWs of baryonic matter blown off the PNS surface occur in low-mass CCSN explosions in analogy to 1D models. In contrast, the innermost ejecta of CCSNe of more massive progenitors comprise the neutrino-heated matter of the inflow-outflow accretion cycle. This highly turbulent environment with secondary shocks provides nucleosynthesis conditions significantly different from the quasi-spherical NDWs.
\item New-born NSs receive natal recoil kicks by anisotropic neutrino emission, which dominates in low-mass CCSN explosions, and by asymmetric mass ejection, which is the main effect in CCSNe of more massive progenitors and can accelerate the NSs for 10\,s and longer. Asymmetric accretion of fallback matter on even longer time scales might transfer considerable amounts of angular momentum to the NS, which could override the spin-up of the PNS caused by asymmetric mass accretion during the onset and early phase of the SN explosion.
\item Neutrino signals from state-of-the-art SN models are basically consistent with the general properties of the detected SN~1987A neutrinos, also during the accretion phase of the first second after bounce. However, the latest three Kamiokande~II events are incompatible with the short cooling times of convective PNSs.
\item These last three Kamiokande~II events might point to less strong and shorter PNS convection than obtained with current nuclear EoSs, or to a seconds-long episode of massive fallback and accretion emission by the NS, or neutrino emission connected to late energy release from a nuclear phase transition. Since the hot PNS EoS has also a strong influence on the success or failure of the neutrino-driven mechanism and the explosion properties of SNe, the PNS EoS remains one of the major uncertain ingredients in CCSN modeling.
\item Stellar-mass BH formation is possible in collapsing stars on different time scales and accompanied by SN explosions of different strengths. The corresponding formation ``channels'' depend on the mass accretion rate by the PNS, its EoS, and the power of neutrino-energy transfer to the medium surrounding the PNS. The associated progenitor masses are unclear because of stellar-evolution uncertainties and disagreements in stellar CC outcomes with different simulation codes.
\item Relevant aspects of the long-time evolution of PNSs and of the SN explosion, although involving 3D physics, can still be well treated in 1D. This holds, for example, for convection in the PNS interior, where MLT yields a good match of the structure and neutrino emission of MD PNS models. Making use of such possibilities in the \textsc{Nemesis} code, the neutrino effects on the long-time SN hydrodynamics can be handled with an enormous gain of computational efficiency. 
\end{enumerate}
\end{summary}

\begin{issues}[FUTURE ISSUES]
\begin{enumerate}
\item  The discrepant results for the explodability of progenitors and for the SN properties obtained with different numerical codes in 3D demand a detailed and well controlled code comparison, testing different setups, resolutions, and physics inputs in particular for the crucial neutrino rates. Such a task, however, will be challenging because of its high demands on computational resources and work force.
\item In order to account for the macroscopic consequences of collective fast neutrino flavor conversion in global, long-term 3D CCSN models, computationally efficient treatments are needed that effectively describe the neutrino flavor evolution obtained by direct solutions of the quantum-kinetic problem, because full-scale SN models cannot resolve the relevant small time and length scales and also lack accurate information on the neutrino angular distributions as long as 3D Boltzmann transport with high resolution is not feasible in these models.  
\item Despite a growing body of work on CCSN modeling with rotation and magnetic fields, the role of these aspects needs to be better understood, in particular by replacing constructed pre-collapse conditions by stellar progenitor models that took into account rotation and magnetic fields consistently over long periods of evolution.
\item The far majority of MD CCSN simulations, in particular those for larger model grids and following the long-term evolution, has employed progenitors from a single stellar evolution code (\textsc{Kepler}; \citenum{Sukhbold+2018}). Because of significant differences in the progenitor structures and their variation with ZAMS mass, 3D CCSN modeling should also be extended to sets of pre-collapse models obtained with other stellar evolution codes.
\item The nuclear EoS of hot PNS matter and a consistent treatment of the neutrino reactions are of crucial importance for predicting the remnant and explosion properties of CCSNe. Systematic studies with different available EoSs that are compatible with all theoretical, experimental, and astrophysical constraints are needed, including tests of the consequences of well motivated phase-transition scenarios.
\item Reliable MD CCSN simulations, strengthened by their scrutinized compatibility with observational SN properties, will permit stronger or more solid bounds on beyond-standard-model particle physics by including such effects into self-consistent SN models.
\item From the astronomical and astrophysical perspective, including GW astronomy, large sets of ultimately 3D CC models for different kinds of progenitors including single stars and binary evolution outcomes and different metallicities are needed to answer the pressing questions connected to observed properties of different types of CCSNe and their compact and gas remnants, of SN nucleosynthesis and galactic chemical evolution, and of GW sources whose statistics will tremendously grow over time with further improved detection sensitivity and planned future instruments.
\item Current self-consistent 3D CCSN simulations need to be extended to much later times, well beyond shock breakout from the stars, to put their predictions to a test by detailed comparison with observations of SNe and SN remnant properties. Multimessenger signals (neutrinos, GWs, and electromagnetic radiation in many wavelength bands) from the next galactic SN (or SNe?) will provide an ultimate test for the theory of neutrino-driven explosions. 
\end{enumerate}
\end{issues}

\section*{DISCLOSURE STATEMENT}
The author is not aware of any affiliations, memberships, funding, or financial holdings that might be perceived as affecting the objectivity of this review.

\section*{ACKNOWLEDGMENTS}
The author is very grateful to Robert Bollig, Robert Glas, Malte Heinlein, Daniel Kresse, Alessandro Lella, and Bernhard M\"uller for their inputs, and to Daniel Kresse also for a careful reading of the manuscript. The work was supported by the German Research Foundation (DFG) through the Collaborative Research Centre ``Neutrinos and Dark Matter in Astro- and Particle Physics (NDM),'' Grant No.\ SFB-1258-283604770, and under Germany's Excellence Strategy through the Cluster of Excellence ORIGINS EXC-2094-390783311. Computing resources are acknowledged from the Max Planck Computing and Data Facility (MPCDF) on the HPC systems Cobra, Draco, and Raven, from the Gauss Centre for Supercomputing e.V.\ (GCS; www.gauss-centre.eu) under GAUSS Call~13 project ID pr48ra, GAUSS Call~15 project ID pr74de, and GAUSS Call~17 and Call~20 project ID pr53yi, and from the Leibniz Supercomputing Centre (LRZ; www.lrz.de) on SuperMUC and SuperMUC-NG at LRZ under LRZ project IDs pn69ho and pn25me.

\bibliography{sn-bibliography-ARNPS}{}
\bibliographystyle{ar-style5.bst}

\end{document}